\documentstyle[12pt]{article}
\def\journal{\topmargin .3in	\oddsidemargin .5in
	\headheight 0pt	\headsep 0pt
	\textwidth 5.625in % 1.2 preprint size  %6.5in
	\textheight 8.25in % 1.2 preprint size 9in
	\marginparwidth 1.5in
	\parindent 2em
	\parskip .5ex plus .1ex		\jot = 1.5ex}
\journal

\catcode`\@=11
\def\marginnote#1{}
\catcode`\@=11
\def\section{\@startsection {section}{1}{0pt}{-3.5ex plus -1ex minus
 -.2ex}{2.3ex plus .2ex}{\raggedright\large\bf}}
\catcode`\@=12
\newskip\humongous \humongous=0pt plus 1000pt minus 1000pt
\def\caja{\mathsurround=0pt}
\def\eqalign#1{\,\vcenter{\openup1\jot \caja
	\ialign{\strut \hfil$\displaystyle{##}$&$
	\displaystyle{{}##}$\hfil\crcr#1\crcr}}\,}
\newif\ifdtup
\def\panorama{\global\dtuptrue \openup1\jot \caja
	\everycr{\noalign{\ifdtup \global\dtupfalse
	\vskip-\lineskiplimit \vskip\normallineskiplimit
	\else \penalty\interdisplaylinepenalty \fi}}}
\def\eqalignno#1{\panorama \tabskip=\humongous
	\halign to\displaywidth{\hfil$\displaystyle{##}$
	\tabskip=0pt&$\displaystyle{{}##}$\hfil
	\tabskip=\humongous&\llap{$##$}\tabskip=0pt
	\crcr#1\crcr}}
\def\R{{\rm I\!R}}

\def\one{{\mathchoice {\rm 1\mskip-4mu l} {\rm 1\mskip-4mu}
{\rm 1\mskip-4.5mu l} {\rm 1\mskip-5mu l}}}
\def\Q{{\mathchoice
{\setbox0=\hbox{$\displaystyle\rm Q$}\hbox{\raise 0.15\ht0\hbox to0pt
{\kern0.4\wd0\vrule height0.8\ht0\hss}\box0}}
{\setbox0=\hbox{$\textstyle\rm Q$}\hbox{\raise 0.15\ht0\hbox to0pt
{\kern0.4\wd0\vrule height0.8\ht0\hss}\box0}}
{\setbox0=\hbox{$\scriptstyle\rm Q$}\hbox{\raise 0.15\ht0\hbox to0pt
{\kern0.4\wd0\vrule height0.7\ht0\hss}\box0}}
{\setbox0=\hbox{$\scriptscriptstyle\rm Q$}\hbox{\raise 0.15\ht0\hbox to0pt
{\kern0.4\wd0\vrule height0.7\ht0\hss}\box0}}}}
\def\C{{\mathchoice
{\setbox0=\hbox{$\displaystyle\rm C$}\hbox{\hbox to0pt
{\kern0.4\wd0\vrule height0.9\ht0\hss}\box0}}
{\setbox0=\hbox{$\textstyle\rm C$}\hbox{\hbox to0pt
{\kern0.4\wd0\vrule height0.9\ht0\hss}\box0}}
{\setbox0=\hbox{$\scriptstyle\rm C$}\hbox{\hbox to0pt
{\kern0.4\wd0\vrule height0.9\ht0\hss}\box0}}
{\setbox0=\hbox{$\scriptscriptstyle\rm C$}\hbox{\hbox to0pt
{\kern0.4\wd0\vrule height0.9\ht0\hss}\box0}}}}

\font\fivesans=cmss10 at 4.61pt
\font\sevensans=cmss10 at 6.81pt
\font\tensans=cmss10
\newfam\sansfam
\textfont\sansfam=\tensans\scriptfont\sansfam=\sevensans\scriptscriptfont
\sansfam=\fivesans
\def\sans{\fam\sansfam\tensans}
\def\Z{{\mathchoice
{\hbox{$\sans\textstyle Z\kern-0.4em Z$}}
{\hbox{$\sans\textstyle Z\kern-0.4em Z$}}
{\hbox{$\sans\scriptstyle Z\kern-0.3em Z$}}
{\hbox{$\sans\scriptscriptstyle Z\kern-0.2em Z$}}}}

\mathchardef\endbar="375

\def\ceilfill{$\raise3pt\hbox{$\mathsurround=0pt\mathord\endbar$}
  \mkern-2mu \xleaders\hbox{$\mkern-5mu
  \mathord-\mkern-5mu$}\hfill\mkern-7mu
  \raise3pt\hbox{$\mathsurround=0pt\mathord\endbar$}$}

\def\floorfill{$\raise9pt\hbox{$\mathsurround=0pt\mathord\endbar$}
  \mkern-2mu \xleaders\hbox{$\mkern-5mu
  \mathord-\mkern-5mu$}\hfill\mkern-7mu
  \raise9pt\hbox{$\mathsurround=0pt\mathord\endbar$}$}

\def\overcontract#1{\mathop{\vbox{\ialign{##\crcr\noalign{\kern3pt}
  \ceilfill\hskip6pt\crcr\noalign{\kern3pt\nointerlineskip}
  $\hfil\displaystyle{#1}\hfil$\crcr}}}}

\def\undercontract#1{\mathop{\vtop{\ialign{##\crcr
  $\hfil\displaystyle{#1}\hfil$\crcr\noalign{\kern3pt\nointerlineskip}
  \floorfill\hskip6pt\crcr\noalign{\kern3pt}}}}}
\def\a{\alpha}
\def\b{\beta}
\def\g{\gamma}
\def\d{\delta}

\def\m{\mu}
\def\n{\nu}

\def\p{\pi}

\def\r{\rho}
\def\s{\sigma}

\def\o{\omega}

\def\x{\xi}
\def\et{\eta}

\def\D{\Delta}
\def\bz{\bar{z}}
\def\bc{\bar{\g}}

\def\bu{\bar{u}}

\def\bq{\bar{q}}

\def\br{\bar{r}}

\def\bA{\bar{A}}

\def\bY{\bar{Y}}
\def\Fb{\bar{F}}

\def\bW{\bar{W}}
\def\tL{\tilde{L}}
\def\tT{\tilde{T}}
\def\tD{\tilde{\Delta}}
\def\P{{\cal P}}

\def\cA{{\cal A}}

\def\cD{{\cal D}}
\def\cbD{\bar{{\cal D}}}
\def\T{{\cal T}}
\def\C{{\cal C}}
\def\cO{{\cal O}}

\def\pa{\partial}
\def\bpa{\bar{\partial}}
\def\ve{\vert}

\def\ra{\rightarrow}
\def\lra{\leftrightarrow}
\def\ti{\times}
\def\oti{\otimes}

\def\xx{\hbox{ }^*_*}
\def\ep{\hfill $\Box$}
\def\ch#1#2{\left( #1 \atop #2 \right)}
\def\cl#1#2#3#4#5#6{\left( \matrix{ #1 & #2 & #3 \cr #4 & #5 & #6 \cr} \right)}
\def\un{\underline}
\def\ni{\noindent}
\def\nl{\newline}

\begin{document}
\begin{titlepage}
\begin{center}
 December 1993     \hfill        UCB-PTH-93/33 \\
 hep-th/9312050  \hfill     LBL-34901 \\
 \hfill  CPTH-A277.1293
%\vskip .05in

{\large \bf Flat Connections and Non-Local Conserved Quantities \\
 in Irrational Conformal Field Theory }
\footnote{The work of MBH was supported in part by the Director, Office of
Energy Research, Office of High Energy and Nuclear Physics, Division of
High Energy Physics of the U.S. Department of Energy under Contract
DE-AC03-76SF00098 and in part by the National Science Foundation under
grant PHY90-21139. The work of NO was supported in part by a
S\'ejour Scientifique de Longue Dur\'ee of the Minist\`ere des Affaires
Etrang\`eres.}

\vskip .1in
M.B. Halpern
\footnote{e-mail: MBHALPERN@LBL.GOV, THEORY::HALPERN}
\vskip .05in
{\em  Department of Physics, University of California\\
      Theoretical Physics Group, Lawrence Berkeley Laboratory\\
      Berkeley, California 94720 \\
       USA}
\vskip .03in
% and
\vskip .1in
N.A. Obers
\footnote{e-mail: OBERS@ORPHEE.POLYTECHNIQUE.FR }
\vskip .03in
{\em Centre de Physique Th\'eorique \\
Ecole Polytechnique  \\
F-91128 Palaiseau \\
France}

\end{center}
\vskip .1in
\begin{abstract}
Irrational conformal field theory (ICFT) includes rational conformal
field theory as a small subspace, and the affine-Virasoro Ward identities
describe the biconformal correlators of ICFT.
We reformulate the Ward identities as an
equivalent linear partial differential system with flat connections and new
non-local conserved quantities. As examples of the formulation, we solve the
system of flat connections for
the coset correlators, the correlators of the affine-Sugawara nests and the
high-level $n$-point correlators of ICFT.
\end{abstract}
\end{titlepage}
\renewcommand{\thepage}{\arabic{page}}
\setcounter{page}{1}
\setcounter{footnote}{1}
\section{Introduction}

Affine Lie algebra, or current algebra on the circle, was discovered
independently in mathematics \cite{KM} and physics \cite{BH}.

It is now understood that
affine Lie algebra underlies rational conformal field theory (RCFT) and
irrational conformal field theory (ICFT),
which includes RCFT as a small subspace,
$$
{\rm ICFT} \supset {\rm RCFT}\;\;\;\;.
\eqno(1.1) $$
The chiral stress tensors of
ICFT live in the universal enveloping algebra of the affine algebra, where they
are given by the general affine-Virasoro construction [3,4],
$$
T = L^{ab} \xx J_a J_b \xx
\eqno(1.2)$$
where $J_a,\,a=1,\ldots,{\rm dim}\,g$ are the currents of affine $g$ and
$L^{ab}$ is a solution of the Virasoro master equation [3,4].

The solution space
of the master equation includes the affine-Sugawara constructions
[2,5-7] the coset constructions [2,5,8], the affine-Sugawara nests
[9-11]
and a vast number of new conformal field theories with generically
irrational central charge. Partial classification of the solution space and
other developments in the Virasoro master equation are reviewed in \cite{SB}.

Recently, the present authors reported a system of equations, the
affine-Virasoro Ward identities \cite{WI}, which describe the biconformal
correlators $A(\bz,z)$ of ICFT. As derived, the Ward identities have the form,
$$
\bpa_{j_1} \cdots \bpa_{j_q} \pa_{i_1} \cdots \pa_{i_p} A(\bz,z) \ve_{\bz =z}
=A_g(z) W_{j_1 \ldots j_q,i_1 \ldots i_p} (z) \;\;\;\;,\;\;\;\;
A_g(z) = A(z,z)
\eqno(1.3a)$$
$$
A_g (z) Q_a^g =0 \;\;\;\;,\;\;a =1,\ldots , {\rm dim}\,g
 \eqno(1.3b)$$
where $W_{j_1 \ldots j_q,i_1 \ldots i_p}$ are the  affine-Virasoro
connections and $A_g$ is the affine-Suga- \linebreak
wara correlator, which satisfies
the KZ equation \cite{KZ} and the $g$-global Ward identity in (1.3b).

The Ward identities follow from KZ-type null states, which live in the
modules of the
universal enveloping algebra of the affine algebra, and the affine-Virasoro
connections may be computed as averages in the affine-Sugawara construction
on $g$. The system has been completely solved for the coset correlators
[13,14], the correlators of the affine-Sugawara nests \cite{WI2} and the
high-level four-point correlators of ICFT \cite{WI2}. These developments are
reviewed in Ref.\cite{RP}, and relevant results will be quoted as needed in
Section 4.

In this paper, we follow a suggestion of Douglas that the Ward identities
might be cast in a more standard form, as a linear PDE with flat connections.
We find that Douglas' intuition is correct, and  an equivalent
formulation of the Ward identities is the system,
$$
\bpa_i A(\bz,z) = A(\bz,z) \bW_i (\bz, z) \;\;\;\;, \;\;\;\;
\pa_i A(\bz,z) = A(\bz,z) W_i (\bz, z)
\eqno(1.4a)$$
$$
A(\bz,z)\, Q_a(\bz,z) =0 \;\;\;\;, \;\; a=1, \ldots ,{\rm dim}\, g
 \eqno(1.4b)$$
where $\bW_i,\,W_i$ are the flat connections of ICFT. The flat connections
are non-linear functionals of the affine-Virasoro connections, whose form
we shall obtain in the development below.

Perhaps more surprising is the existence of the non-local conserved generators
$Q_a(\bz,z)$ of Lie $g$,
$$
[ Q_a(\bz,z) , Q_b (\bz,z) ] =i f_{ab}{}^c Q_c(\bz,z)
\eqno(1.5)$$
which appear in (1.4b). These generators, which are also non-linear
functionals of the affine-Virasoro connections, are the  lift of
the global generators $Q_a^g$ into the space of ICFT's: Comparing
eqs.(1.3b) and (1.4b), we see that the non-local generators play the
same role in the general theory - restriction of the correlators to the singlet
sector of $g$ - that the global generators play in the affine-Sugawara
construction. Moreover, we shall see that the non-local generators include the
usual global generators of $h$,
$$
Q_a (\bz,z) = Q_a^g \;\;\;\;, \;\; a \in h
\eqno(1.6)$$
 when $h \subset g$ is an ordinary symmetry of
the construction.

The equivalence of the Ward identities (1.3) and the system of flat
connections (1.4) is established in Sections 2 and 3. In Section 4, we
begin to  translate what is known about the affine-Virasoro
connections into the language of flat connections, focusing on
$SL(2,\R)$ covariance, K-conjugation covariance, crossing symmetry and the
high-level form of the flat connections on simple $g$.
Using some of these properties, we also obtain the flat connections of the
cosets and nests in Section 5.

In Section 6, we discuss our results in further detail for the coset
correlators \cite{WI}, obtained as usual by factorizing the biconformal
correlators
of $h$ and $g/h$. In this simple case, we see explicitly that flat connections
and non-local conserved quantities are induced for the conformal correlators
as well.

Finally, Sections 7 and 8 solve the high-level system of flat connections to
obtain the high-level $n$-point biconformal and conformal correlators of
ICFT.
The results for the correlators exhibit induced flat connections and non-local
conserved quantities for all ICFT, agree with the known form of the high-level
four-point correlators \cite{WI2} and continue to show good physical properties
when
$n \neq 4$.

\section{Flat Connections and Ward Identities}

In this section, we study a particular system of linear partial differential
equations,
$$\bpa_i A(\bz,z) = A(\bz,z) \bW_i(\bz,z)
\;\;\;\;,\;\;\;\;  \pa_i A(\bz,z) = A(\bz,z) W_i(\bz,z)
 \eqno(2.1a)$$
$$ (\bz,z) = \{ \bz_1,\ldots,\bz_n, z_1 , \ldots, z_n \}
\eqno(2.1b)$$
$$ W_i^g(z) \equiv \bW_i(z,z)+W_i (z,z)
\eqno(2.1c)$$
under the assumption of consistency. Here $A^{\a}$, $\a =1,\ldots ,D$ is a row
vector, $(\bW_i)_{\a}{}^{\b}$ and $(W_i)_{\a}{}^{\b}$ are
$D \ti D$ matrices, and the $\bz$'s are not
necessarily complex conjugates of the $z$'s. In what follows, we refer to
$A$ and $\bW_i,\;W_i$ as the {\it bicorrelator} and its {\it connections}
respectively.

Note that the system (2.1) is not the most general PDE on $2n$ variables.
Implicit in the definition (2.1c) is the restriction that the pinched
flat connections
$$
\bW_i (z,z) = \bW_i(\bz,z) \ve_{\bz =z } \;\;\;\;,\;\;\;\;
W_i(z,z) = W_i (\bz,z) \ve_{\bz = z}
\eqno(2.2)$$
are finite, where $\bz =z$ means $\bz_i = z_i$ for all $i = 1,\ldots , n$.
In this way, we exclude, for example, the KZ equation \cite{KZ} for a
$2n$-point
correlator, whose connections are singular on this line. In what follows,
we refer to the quantity $W_i^g$ in (2.1c) as the {\it base connection}
of the system.

\ni \un{Proposition 2.1}.The connections  satisfy the
flatness conditions,
$$ \bpa_i \bW_j - \bpa_j \bW_i + [\bW_i,\bW_j] =0
\eqno(2.3a)$$
$$ \pa_i W_j - \pa_j W_i + [W_i,W_j] =0
\eqno(2.3b)$$
$$ (\pa_i + W_i) \bW_j = (\bpa_j + \bW_j) W_i
\eqno(2.3c)$$
which say that $\bW_i$ and $W_i$ are flat, and satisfy the cross-flatness
condition (2.3c). \nl
\ni \un{Proof}. These relations follow as usual, by differentiation
of (2.1a).
\ep

\ni \un{Definition 2.2}. The differential operators
$$
  \cbD_i (\bz,z) = \bpa_i + \bW_i(\bz,z) \;\;\;\;,\;\;\;\;
\cD_i (\bz,z) = \pa_i + W_i(\bz,z)
\eqno(2.4)$$
are called the {\it covariant derivatives}.

\ni \un{Proposition 2.3}. All covariant derivatives commute,
$$ [\cbD_i,\cbD_j ] = [\cD_i ,\cD_j] = [\cD_i, \cbD_j] =0
\;\;\;\;.
\eqno(2.5)$$
\ni \un{Proof}. The commutativity follows immediately from Proposition 2.1.
\ep

\ni \un{Definition 2.4}. The non-linear functionals of the flat connections,
$$ W_{j_1\ldots j_q,i_1\ldots i_p} \equiv \cbD_{j_1} \cdots \cbD_{j_q}
\cD_{i_1} \cdots \cD_{i_p} \one \ve_{\bz = z}
 \eqno(2.6)$$
are called the {\it connection moments}, where $\one$ is the $D \ti D$ unit
matrix.
As  examples, we have
$$
W_{0,0} = \one \;\;\;\;,\;\;\;\; W_{i,0}(z) = \bW_i (z,z) \;\;\;\;,\;\;\;\;
W_{0,i} (z) = W_i (z,z)
\eqno(2.7)$$
where $\bW_i (z,z) $ and $W_i (z,z)$ are the pinched flat connections in
(2.2).

\ni \un{Proposition 2.5}. The connection moments are symmetric under exchange
of
any pair of
$i$'s or any pair of $j$'s. \nl
\ni \un{Proof}. The symmetry follows immediately from Proposition 2.3 and
Definition 2.4.
\ep

\ni \un{Proposition 2.6}. The base connection is a flat connection,
$$ \pa_i W_j^g - \pa_j W_i^g + [W_i^g,W_j^g] = 0 \;\;\;\;.
\eqno(2.8)$$
\ni \un {Proof}. Use (2.1c), Proposition 2.1 and the differentiation rule
$$ \pa_i [f(\bz,z)  \ve_{\bz  = z}] =
[ (\pa_i + \bpa_i) f(\bz,z)] \ve_{\bz =z} \;\;\;\;.
\eqno(2.9)$$
\ep

\ni \un{Proposition 2.7}. The connection moments satisfy the {\it consistency
relations},
$$ (\pa_i + W_i^g) W_{j_1\ldots j_q, i_1 \ldots i_p} =
 W_{j_1 \ldots j_q i , i_1 \ldots i_p} + W_{j_1 \ldots j_q,i_1 \ldots i_pi}
\;\;\;\;.
\eqno(2.10)$$
\ni \un{Proof}. Use (2.1c), (2.9), Definition 2.2,
Definition 2.4 and Proposition
2.3 to verify
$$
\eqalign{
{\rm l.h.s.} = (\bpa_i + \pa_i + & \bW_i  + W_i )
\cbD_{j_1} \cdots \cbD_{j_q}
\cD_{i_1} \cdots \cD_{i_p} \one \ve_{\bz = z} \cr
= & ( \cbD_i + \cD_i) \cbD_{j_1} \cdots \cbD_{j_q}
\cD_{i_1} \cdots \cD_{i_p} \one \ve_{\bz = z} = {\rm r.h.s.}
\;\;\;\;. \cr}
\eqno(2.11)$$
\ep
\nl \ni \un{Remark}. The consistency relations (2.10) were given in
eq.(8.11) of Ref.\cite{WI}. In eq.(4.1) of Ref.\cite{WI2}, these relations are
solved to
express all the connection moments in terms of the base connection and
the {\it one-sided} connection
moments $W_{j_1 \ldots j_q, 0}$ or $W_{0,i_1 \ldots i_p}$.

\ni \un{Definition 2.8}. The {\it base correlator} $A_g^{\a},\;
\a =1 , \ldots , D$ is defined as
$$ A_g(z) \equiv A(z,z) \;\;\;\;. \eqno(2.12)$$

\ni \un{Proposition 2.9}. The base correlator satisfies the {\it base
equation},
$$ \pa_i A_g (z) = A_g(z) W_i^g (z)
\;\;\;\;.  \eqno(2.13)$$
\ni \un{Proof}. Differentiate the base correlator with eq.(2.9), and
use eq.(2.1).
\ep

\ni \un{Proposition 2.10}. The bicorrelator satisfies the {\it Ward
identities},
$$
\bpa_{j_1} \cdots \bpa_{j_q} \pa_{i_1} \cdots \pa_{i_p} A(\bz,z)
\ve_{\bz = z} = A_g(z) W_{j_1 \ldots j_q, i_1 \ldots i_p} (z)
\eqno(2.14a)$$
$$
A_g(z) = A(z,z) \;\;\;\;,\;\;\;\; \pa_i A_g(z) = A_g(z) W_i^g (z)
\;\;\;\;. \eqno(2.14b)$$
\ni \un{Proof}. Differentiate $A$ as indicated, using (2.1) and
Definitions 2.4 and 2.8.
\ep

\ni \un{Definition 2.11}. The (invertible) {\it evolution operators}
$\Fb_{\a}{}^{\b}$ and $F_{\a}{}^{\b}$ of the
flat connections in  (2.1a) are
$$
\Fb(\bz,z) \equiv P^* \mbox{e}^{\int_{z}^{\bz} {\rm d} \bz_i ' \bW_i(\bz ',z) }
\eqno(2.15a)$$
$$
F(\bz,z) \equiv P^* \mbox{e}^{\int_{\bz}^z {\rm d} z_i' W_i(\bz,z') }
\eqno(2.15b)$$
where $P^*$ is anti-path ordering. These satisfy
$$ \bpa_i \Fb = \Fb \bW_i \;\;\;\;,\;\;\;\; \pa_i F = F W_i
\eqno(2.16a)$$
$$ \Fb(z,z) = F(z,z) = \one \;\;\;\;.
 \eqno(2.16b)$$

\ni \un{Proposition 2.12}. The evolution operators may be written as
$$ \Fb(\bz,z) = \sum_{q=0}^{\infty} {1 \over q !} \sum_{j_1 \ldots j_q}
\prod_{\n=1}^q (\bz_{j_\n} - z_{j_\n } ) W_{j_1 \ldots j_q,0} (z)
\eqno(2.17a)$$
$$ F(\bz,z) = \sum_{p=0}^{\infty} {1 \over p !} \sum_{i_i \ldots i_p}
\prod_{\m=1}^p (z_{i_\m} - \bz_{i_\m } ) W_{0,i_1 \ldots i_p} (\bz)
\eqno(2.17b)$$
where $W_{0,i_1 \ldots i_p}$ and $W_{j_1 \ldots j_q,0} $ are the one-sided
connection moments. \nl
\ni \un{Proof}. To obtain (2.17b), follow the steps,
$$ \eqalign{ F(\bz,z) & = F (\bz, \bz + z -\bz) \cr
& = \sum_{p=0}^{\infty} {1 \over p!} \sum_{i_1 \ldots i_p}
\prod_{\m=1}^p (z_{i_\m} - \bz_{i_\m } ) \pa_{i_1} \cdots \pa_{i_p} F(\bz,z)
\ve_{z = \bz} \cr
& = \sum_{p=0}^{\infty} {1 \over p!} \sum_{i_1 \ldots i_p}
\prod_{\m=1}^p (z_{i_\m} - \bz_{i_\m } ) \cD_{i_1} \cdots \cD_{i_p} \one
\ve_{z = \bz} \cr
& = \sum_{p=0}^{\infty} {1 \over p!} \sum_{i_1 \ldots i_p}
\prod_{\m=1}^p (z_{i_\m} - \bz_{i_\m } ) W_{0,i_1 \ldots i_p} (\bz) \cr}
 \eqno(2.18)$$
where (2.16) and Definition 2.4 were used
in the last two lines. The result in (2.17a) follows similarly from
$\Fb (\bz,z) = F (z + \bz -z ,z) $.
\ep

\ni \un{Proposition 2.13}. The bicorrelator can be expressed in terms of the
base correlator and the evolution operators,
$$ A(\bz,z) = A_g (z) \Fb(\bz,z) = A_g (\bz) F(\bz,z)
\;\;\;\;.
\eqno(2.19)$$
\ni \un{Proof}. These relations follow immediately from Definition 2.8 and the
properties
(2.16)  of  $\Fb,\; F$.
\ep

\ni \un{Definition 2.14}. The (invertible) evolution operator of the
base connection,
$$
A_g(z,z_0) \equiv P^* \mbox{e}^{\int_{z_0}^{z} {\rm d} z_i ' W_i^g (z ') }
\eqno(2.20)$$
will be called the {\it base evolution operator}. Some well-known properties of
the
base evolution operator include
$$ \pa_i A_g(z,z_0) = A_g(z,z_0) W_i^g (z) \;\;\;\;,\;\;\;\;
\pa_i^0 A_g(z,z_0) = - W_i^g (z_0) A_g(z,z_0) \eqno(2.21a)$$
$$
A_g(z_0,z_0) = \one
\;\;\;\;,\;\;\;\;
A_g^{-1} (z,z_0) = A_g (z_0 ,z)
\eqno(2.21b)$$
$$ A_g(z_0,z) A_g (\bz,z_0) = A_g(\bz,z)
\eqno(2.21c)$$
$$
A_g^{\a}(z)=A_g(z_0)^{\b}  A_g(z,z_0)_{\b}{}^{\a} \;\;\;\; .
\eqno(2.21c)$$

\ni \un{Proposition 2.15}. The evolution operators $\Fb$ and $F$ are related
by the base evolution operator,
$$ \Fb (\bz,z) = A_g (\bz,z) F(\bz,z)
\;\;\;\;, \;\;\;\;  F(\bz,z) = A_g (z,\bz) \Fb (\bz ,z )  \;\;\;\;.
\eqno(2.22)$$
\ni \un{Proof}. Use Proposition 2.13 and (2.21b-d).

\ni \un{Proposition 2.16}. The
evolution operators $\Fb$ and $F$ satisfy
$$ (\pa_i + W_i^g (z) ) \Fb (\bz,z)
= \Fb(\bz,z) W_i (z)
\eqno(2.23a) $$
$$
\;\;\;\; (\bpa_i + W_i^g (\bz) ) F(\bz,z)  = F(\bz,z) \bW_i (\bz)
\;\;\;\;. \eqno(2.23b)$$
\ni \un{Proof}. To obtain (2.23a),
use (2.21a) and (2.16a) to
differentiate the first relation in (2.22). To obtain
(2.23b), differentiate the second relation in (2.22).
\ep

\ni \un{Proposition 2.17}. The flat connections can be expressed as non-linear
functionals of
the connection moments,
$$ \bW_i = \Fb^{-1} \bpa_i \Fb = F^{-1} (\bpa_i + W_i^g (\bz) ) F
\eqno(2.24a)$$
$$ \;\;\;\;\;
W_i = F^{-1} \pa_i F = \Fb^{-1} (\pa_i + W_i^g (z) ) \Fb \;\;\;\;.
\eqno(2.24b)$$
\ni \un{Proof}. Because $\Fb$ and $F$ are invertible, these forms follow
immediately from Definition 2.11 and Proposition  2.16.
\ep \nl
\un{Remark}. The  relations (2.24) are the inverse
of eq.(2.6), which expresses the connection moments in terms of the
connections. See Appendix B for further details of the inversion
(2.24).

Note also that the first forms in
(2.24) give $\bW_i$ and $W_i$ as  manifestly  flat
connections. The second forms in (2.24) can also be
used to verify the cross-flatness condition (2.3c).

\ni \un{Theorem 2.18}. The system of Ward identities (2.14)
  is equivalent to the
original system (2.1) of flat connections. \nl
\un{Proof}. In the development above, we have shown that the original system
(2.1) implies the Ward identities (2.14).
To see that the Ward identities
imply the original system, Taylor expand the correlators
$$
A(\bz,z) = A(\bz,\bz+z-\bz) = A(z+\bz -z,z)
\eqno(2.25)$$  and use the Ward identities
to obtain eqs.(2.19) in the form,
$$
\eqalign{ \;\;\;\;\;\;\;
A^{\a}(\bz,z) &= A_g^{\b} (\bz) \sum_{p=0}^{\infty} {1 \over p!}
\sum_{i_1 \ldots i_p} \prod_{\m = 1}^p (z_{i_{\m}}-\bz_{i_{\m}})
{W_{0,i_1 \ldots i_p} (\bz)_{\b}}^{\a}  \cr
 & = A_g^{\b} (z) \sum_{q=0}^{\infty} {1 \over q!}
\sum_{j_1 \ldots j_q} \prod_{\n =1}^q (\bz_{j_{\n}}-z_{j_{\n}})
{W_{j_1 \ldots j_q,0} (z)_{\b}}^{\a} \;. \;\;\;  \cr }
 \eqno(2.26)$$
Then we have
$$
\bpa_i A(\bz,z) = A_g(z) \bpa_i \Fb (\bz,z) =  A(\bz,z) \bW_i (\bz,z)
\eqno(2.27a)$$
$$
\pa_i A(\bz,z) = A_g(\bz) \pa_i F(\bz,z) =  A(\bz,z) W_i (\bz,z)
\eqno(2.27b)$$
by equation (2.16a).
\ep \nl \un{Remark}. The relations (2.26) were given in eq.(6.13) of
Ref.\cite{WI}.

\section{Non-Local Conserved Quantities}

In this section, we introduce some additional structure into the system
(2.1)
of flat connections. In particular, we now assume that the base connection
$W_i^g$ in (2.1c) exhibits a global invariance under a Lie algebra $g$.
\nl
\un{Definition 3.1}. Let $(Q_a^g)_{\a}{}^{\b}$, $a=1,\ldots, {\rm dim}\,g$,
called the {\it global generators of} $g$, be $D \ti D$
representation matrices of Lie $g$, which satisfy
$$ [Q_a^g, W_i^g (z) ] = 0
 \eqno(3.1a)$$
$$ [Q_a^g , Q_b^g ] = i f_{ab}{}^c Q_c^g \;\;\;\;,\;\;\;\;
\bpa_i Q_a^g = \pa_i Q_a^g = 0  \eqno(3.1b)$$
where $ f_{ab}{}^c$ are the structure constants of $g$.

\ni \un{Proposition 3.2}. The base evolution operator in
(2.20) is $g$-invariant,
$$
[ Q_a^g , A_g(z,z_0) ] = 0
\;\;\;\;.
\eqno(3.2)$$
\un{Proof}. The commutativity follows immediately from (3.1a) and
(2.20).
\ep

\ni \un{Proposition 3.3}. The quantities $(A_g\, Q_a^g)^{\a},\;a=1, \ldots ,
 {\rm dim}
\, g $ are covariantly
constant,
$$
\cD_i^g (A_g \, Q_a^g) \equiv \pa_i (A_g \, Q_a^g) - (A_g \, Q_a^g) W_i^g
=0
\eqno(3.3)$$
where $A_g$ is the base correlator in eq.(2.12). \nl
\un{Proof}. Differentiate as indicated, using the base equation (2.13)
and eq.(3.1).
\ep

\ni \un{Proposition 3.4}.
It is consistent to choose the singlet sector of the base correlator,
$$
A_g (z) Q_a^g =0\;\;\;\;,\;\;\;\;a =1 , \ldots , {\rm dim}\,g
\eqno(3.4)$$
which is called the $g$-global Ward identity. In the language of initial value
problems, this means that the choice of a
$g$-invariant initial condition $A_g (z_0) Q_a^g = 0$ is preserved after
evolution to any point $z$,
$$
A_g(z_0) Q_a^g = 0 \;\;\; \ra \;\;\; A_g(z) Q_a^g =0
  \eqno(3.5)$$
and (3.5)
defines the sense in which the global generators are conserved. \nl
\un{Proof}. The statements (3.4) and (3.5) follow immediately from Proposition
3.3.
\nl $\mbox{} \;\, $ \ep  \nl
\un{Remark}. The structure above is well known in the study of
KZ equations \cite{KZ},
where (3.4) is called the $g$-global Ward identity. It is our task to
find the lift of this structure into the system (2.1) of flat connections.

\ni \un{Definition 3.5}. The non-linear functionals of the connection
moments,
$$ Q_a (\bz,z) \equiv F^{-1} (\bz,z) Q_a^g F (\bz,z)
\;\;\;\;,\;\;\;\;a=1,\ldots , {\rm dim}\,g
\eqno(3.6)$$
are called the {\it non-local generators of} $g$, where $F$ is the evolution
operator in (2.15b). The explicit form of these functionals is worked out
in Appendix C.

\ni \un{Proposition 3.6}. The non-local generators are a representation of Lie
$g$,
$$ [Q_a (\bz,z) , Q_b(\bz,z) ] = if_{ab}{}^c Q_c (\bz,z) \;\;\;\;.
\eqno(3.7)$$
\ni \un{Proof}. $Q_a$ is a similarity transformation of $Q_a^g$.
\ep

\ni \un{Proposition 3.7}. The non-local generators have two equivalent forms,
$$  Q_a (\bz,z) = F^{-1} (\bz,z) Q_a^g F (\bz,z)
=\Fb^{-1}(\bz,z) Q_a^g \Fb (\bz,z)
\eqno(3.8)$$
where $\Fb$ is the evolution operator in (2.15a).
\nl \ni \un{Proof}. Using Definition 3.5, Proposition 2.15 and Proposition 3.2,
we have
$$
\eqalign{ \Fb^{-1} (\bz,z) Q_a^g \Fb (\bz,z)
 & = F^{-1}(\bz,z) A_g^{-1}(\bz,z) Q_a^g A_g(\bz,z) F(\bz,z) \cr
 & = F^{-1}(\bz,z)  Q_a^g  F(\bz,z)  \;\;\;\;.  \cr}
\eqno(3.9)$$
\ep

\ni \un{Proposition 3.8}. $Q_a$ is covariantly constant,
 $$
\bar{D}_i Q_a \equiv  \bpa_i Q_a  + [\bW_i,Q_a] = 0
\eqno(3.10a)$$
$$
\;\;\;\;\; D_i Q_a \equiv  \pa_i Q_a  + [W_i,Q_a] = 0 \;\;\;\; .
\eqno(3.10b)$$
\un{Proof}. Differentiate the two forms of $Q_a$ in (3.8), using eqs.
(2.16a) and (3.1b). \nl  $\mbox{} \;$ \ep \nl
\un{Remark}. The relations (3.10) are consistent
because
$W_i$ and $\bW_i$ are
flat and cross-flat.

\ni \un{Proposition 3.9}. The quantities $(A \, Q_a)^{\a},\; a=1,\ldots,{\rm
dim}\,g  $ are covariantly constant,
$$
\bar{D}_i (A \,Q_a) =  \bpa_i (A Q_a ) - (A Q_a) \bW_i = 0
\eqno(3.11a)$$
$$
\;\;\;\;\; D_i (A \,Q_a) =  \pa_i (A Q_a ) - (A Q_a) W_i = 0 \;\;\;\;.
 \eqno(3.11b)$$
\un{Proof}. These relations follow from Proposition 3.8 and
the differential equations (2.1a) for the
bicorrelator.
\ep

\ni \un{Proposition 3.10} It is consistent to choose the singlet sector of the
bicorrelator,
$$
A(\bz,z) Q_a (\bz,z) =0 \;\;\;\;,\;\;\;\;
a= 1, \ldots , {\rm dim}\, g \;\;\;\;.
\eqno(3.12) $$
In the alternate (initial value) language of Proposition 3.4, this means that
the non-local generators of $g$ are conserved quantities,
$$  A( \bz_0,z_0) Q_a (\bz_0,z_0) = 0 \;\;\;
\ra \;\;\; A( \bz,z) Q_a (\bz,z) = 0
\eqno(3.13)$$
under the evolution defined by the system of flat connections.
\nl \un{Proof}. The statements (3.12) and (3.13) follow immediately from
Proposition 3.9.
\ep

\ni \un{Proposition 3.11}. The consistent choice of singlet sectors under
$Q_a^g$
and $Q_a(\bz,z)$ are equivalent,
$$ A_g(z_0) Q_a^g  = 0
\lra A_g(\bz_0,z_0) Q_a (\bz_0,z_0) = 0
\eqno(3.14a)$$
$$ A_g(z) Q_a^g  = 0
\lra A_g(\bz,z) Q_a (\bz,z) = 0 \;\;\;\;.
\eqno(3.14b)$$
\un{Proof}. Using (2.19) and (3.8) we have
$$ A(\bz,z) Q_a (\bz,z) = A_g(z) Q_a^g \Fb (\bz,z) \;\;\;\;.
\eqno(3.15)$$
Because $\Fb$ is invertible, the statements in (3.14) follow in both directions
from
eq.(3.15).
\ep

\ni \un{Theorem 3.12}. Including the conserved quantities, the description
by flat connections is equivalent to the description by Ward identities:
\nl
a) Flat connections.
$$
\bpa_i A (\bz,z) = A (\bz,z)  \bW_i (\bz,z)\;\;\;\;,\;\;\;\;
\pa_i A (\bz,z) = A (\bz,z) W_i (\bz,z)
\eqno(3.16a)$$
$$
 A (\bz,z) Q_a (\bz,z) =0 \;\;\;\;,\;\; a=1, \ldots , {\rm dim}\, g
\eqno(3.16b)$$
b) Ward identities.
$$
\bpa_{j_1} \cdots \bpa_{j_q} \pa_{i_1} \cdots \pa_{i_p} A(\bz,z)
\ve_{\bz = z} = A_g(z) W_{j_1 \ldots j_q, i_1 \ldots i_p} (z)
\;\;\;\;,\;\;\;\;
A_g(z) = A(z,z) \eqno(3.17a)$$
$$ A_g(z) Q_a^g = 0 \;\;\;\;, \;\; a= 1,\ldots , {\rm dim}\, g
\;\;\;\; .
\eqno(3.17b)$$
\un{Proof}. The equivalence of the differential systems
(3.16a) and (3.17a) was established in Theorem 2.18. The equivalence of
the statements of $g$-invariance in (3.16b) and (3.17b)
summarizes the conclusions of this section.
\ep

\section{Irrational Conformal Field Theory}

We have shown a 1-1 correspondence between the Ward identities
(3.17) and the
system (3.16a) of flat connections including
the non-local conserved generators of $g$
in (3.16b).

As noted in the introduction, the affine-Virasoro Ward identities
\cite{WI} have
exactly the
form (3.17),
and so irrational conformal field theory (ICFT) provides us with a very large
class of concrete realizations of the
system of flat connections\footnote{The invariant four-point Ward identities
\cite{WI} are also equivalent to a system of invariant flat connections (see
Appendix A).}.
In these realizations, we have the following translation
dictionary.

\ni A.
$ \mbox{Bicorrelators} \rightarrow \mbox{biconformal correlators} $. \nl
Biconformal correlators are  averages of Virasoro biprimary fields
 [16,13], which are simultaneously Virasoro primary under the commuting
K-conjugate pair [2,5,8,3] of affine-Virasoro constructions,
$$
\tT = \tL^{ab} \xx J_a J_ b \xx \;\;\;\;, \;\;\;\;
T = L^{ab} \xx J_a J_b \xx
\eqno(4.1a)$$
$$
\tL^{ab} + L^{ab} = L_g^{ab} \;\;\;\;,\;\;\;\;
\tT + T = T_g = L^{ab}_g \xx J_a J_b \xx \;\;\;\;.
\eqno(4.1b)$$
Here, $J_a,\; a=  1,\ldots, {\rm dim}\,g$ are the currents of affine $g$
[2,1], $\tL^{ab}$ and $L^{ab}$ are a K-conjugate pair of solutions to the
Virasoro master equation [3,4], and $T_g$ is the affine-Sugawara construction
on
$g$ [2,5-7].

In this paper, we discuss only the simplest set of biconformal correlators,
$$
 A^{\a}(\bz,z) =  \langle R^{\a_1}(\T^1,\bz_1,z_1) \ldots
R^{\a_n}(\T^n,\bz_n,z_n) \rangle \;\;\;\;,\;\;\;\;\a=(\a_1\ldots\a_n)
 \eqno(4.2a)$$
$$
R^{\a}(\T,\bz,z) = \exp [(\bz-z) \!\oint_z {{\rm d} w \over 2 \pi i } \tT (w) ]
R_g^{\a} (\T,z)
 = \exp [(z-\bz) \! \oint_{\bz}
 {{\rm d} w \over 2 \pi i } T (w) ] R_g^{\a} (\T,\bz)
\eqno(4.2b)$$
because they satisfy the simplest affine-Virasoro Ward identities, given in
(1.3). In this case, we include only the biprimary fields
$R^{\a}(\T,\bz,z),\, \a =1 ,\ldots {\rm dim}\,\T$ of the broken affine-primary
fields, where $R^{\a}_g(\T,z)$ is the
$L^{ab}$-broken affine primary field \cite{WI}
corresponding to  matrix irrep
$\T$ of $g$.

\ni B.
$ \mbox{Base connection}   \ra \mbox{KZ connection on}\;\,g$,  \nl
$ \mbox{ }
\;\;\;\, \mbox{base equation} \ra \mbox{KZ equation on} \;\,g$, \nl
 $ \mbox{ }\;\;\;\,  \mbox{base correlator}  \ra
\mbox{affine-Sugawara correlator on}\;\,g $, \nl
 $\mbox{ }\;\;\;\, \mbox{base evolution operator} \ra
\mbox{evolution operator on}\;\, g$.
\nl The explicit forms of these objects are well known,
$$
 W_i^g(z) = 2 L^{ab}_g \sum_{j\neq i}^n
{\T_a^i \T_b^j \over z_{ij} } \;\;\;\;,\;\;\;\;
Q^g_a = \sum_{i=1}^n \T_a^i  \;\;\;,\;\; a= 1,\ldots,{\rm dim}\, g
\eqno(4.3a)$$
$$
A_g(z) = A(z,z) = \langle R_g (\T^1,z_1) \ldots R_g (\T^n,z_n) \rangle
= A_g(z_0) A_g(z,z_0) \eqno(4.3b)$$
$$
\pa_i A_g (z) = A_g(z)  W_i^g (z) \;\;\;\;, \;\;\;\; A_g (z) Q_a^g = 0
\eqno(4.3c)$$
where $\T_a^i,\; i =1,\ldots, n $ is the matrix irrep of $g$ corresponding
to the $i$th
affine primary field $R_g$.

When $h \subset g$, we also have the flat connections and
correlators of $h$,
$$
W_i^h (z)= 2 L^{ab}_h \sum_{j \neq i} { \T_a^i \T_b^j \over z_{ij} }
\;\;\;\;, \;\;\;\;
\pa_i A_h (z) = A_h (z)  W_i^h (z)
\eqno(4.4a)$$
$$
A_h^{\a} (z) = A_h^{\b} (z_0) A_h(z,z_0)_{\b}{}^{\a}
\eqno(4.4b)$$
$$
 [ Q^g_a,W_i^h (z) ] =[ A_h(z,z_0), Q^g_a] = A_h (z)  Q^g_a = 0
\;\;\;\;,\;\;a \in h
\eqno(4.4c)$$
where $A_h(z,z_0)$ is the (invertible) evolution operator of $h$. The
properties of this evolution operator
mirror those of the evolution operator on $g$, including
$$
\pa_i A_h(z,z_0) = A_h(z,z_0) W_i^h (z) \;\;\;\;,\;\;\;\; A_h(z_0,z_0) = \one
\eqno(4.5a)$$
$$
A_{h} (z,\bz)= A_{h}(z_0,\bz) A_{h}(z,z_0)
= A_{h}^{-1}(\bz,z_0) A_{h}(z,z_0)
\eqno(4.5b)$$
$$
\;\;\;\; A_{h} (\bz,z)= A_{h}(z_0,z) A_{h}(\bz,z_0)
= A_{h}^{-1}(z,z_0) A_{h}(\bz,z_0) \;\;\;\;.
\eqno(4.5c)$$

\ni C.
$ \mbox{Connection moments} \ra \mbox{affine-Virasoro connections} $.
\nl The affine-Virasoro connections
$W_{j_1 \ldots j_q,i_1 \ldots i_p}$ may be evaluated in principle from the
formula [13,14],
$$ A_g^{\b} (z) {W_{j_1 \ldots j_q, i_1 \ldots i_p } (z)_{\b}}^{\a} =
\;\; \;\;\;\;\;\;\;\;\;\;\; \;\;\;\;\;\;\;\;\;\; $$
$$ \left[ \prod_{r=1}^{q} \tL^{a_r b_r} \oint_{z_{j_r}}
{{\rm d}\o_r \over 2\p i} \oint_{\o_r}  {{\rm d}\et_r \over 2\p i} \;
{1 \over \et_r-\o_r} \right] \!
\left[ \prod_{s=1}^{p} L^{c_s d_s} \oint_{z_{i_s}} \!
{{\rm d}\o_{q+s} \over 2\p i} \oint_{\o_{q+s}} \!
{{\rm d}\et_{q+s} \over 2\p i} \; {1 \over \et_{q+s}-\o_{q+s}} \right] $$
$$
\times \langle
J_{a_1}(\et_1)  J_{b_1}(\o_1)  \ldots J_{a_q}(\et_q)  J_{b_q}(\o_q)
J_{c_1}(\et_{q+1})  J_{d_1}(\o_{q+1})  \ldots  $$
$$
 \;\;\;\; \;\;\;\;\;\; \; J_{c_p}(\et_{q+p})  J_{d_p}(\o_{q+p})
R_g^{\a_1} (\T^1,z_1) \ldots  R_g^{\a_n} (\T^n,z_n) \rangle
 \eqno(4.6)$$
since the required averages are in the affine-Sugawara construction on $g$.
Results have been obtained
thru second order
($q+p=2$) for all theories \cite{WI},  to all orders for all coset
constructions  [13,14]
and affine-Sugawara nests \cite{WI2}, and to leading order at high level for
all
ICFT on simple $g$ \cite{WI2}.

\ni D. Formula for the flat connections. \nl
Using (4.1), (4.2) and the system $\bpa_i A = A \bW_i $,
$\pa_i A = A W_i $ of flat connections, we obtain the relations
$$
\eqalign{
 A^{\b} & (\bz,z) \bW_i (\bz,z)_{\b}{}^{\a} \cr
& = \oint_{\bz_i} {{\rm d} w \over 2 \pi i} \oint_w { {\rm d} \eta \over 2\pi
i} {\tL^{ab} \over \eta -w} \langle J_a(\eta) J_b (w)
R^{\a_1} (\T^1,\bz_1,z_1) \ldots R^{\a_n} (\T^n ,\bz_n,z_n) \rangle
\cr }
\eqno(4.7a)$$
$$
\eqalign{
 A^{\b} & (\bz,z) W_i (\bz,z)_{\b}{}^{\a} \cr
& = \oint_{z_i} {{\rm d} w \over 2 \pi i} \oint_w { {\rm d} \eta \over 2\pi i}
{L^{ab} \over \eta -w} \langle J_a(\eta) J_b (w)
R^{\a_1} (\T^1,\bz_1,z_1) \ldots R^{\a_n} (\T^n ,\bz_n,z_n) \rangle
\cr }
\eqno(4.7b)$$
which are equivalent to the formula (4.6) for the one-sided
affine-Virasoro connections. The original form in (4.6) is apparently
superior for computational purposes.

\ni E. $SL(2,\R) \ti SL(2,\R)$ covariance. \nl  Because affine-Virasoro
constructions come in commuting K-conjugate pairs, we know that
the biconformal correlators satisfy the $SL(2,\R) \ti SL(2,\R)$
relations,
$$
\eqalignno{
A \sum_{i=1}^n \bW_i = A \sum_{i=1}^n (\bz_i \bW_i + \tD_i ) & =
A \sum_{i=1}^n (\bz_i^2 \bW_i + 2 \bz_i \tD_i ) = 0
&(4.8a) \cr
 A \sum_{i=1}^n W_i = A \sum_{i=1}^n (z_i W_i + \D_i ) & =
A \sum_{i=1}^n (z_i^2 W_i + 2 z_i \D_i ) = 0 \;\;\;\;.
&(4.8b) \cr} $$
Then, repeated differentiation of (4.8) at $\bz = z$ gives
the corresponding   $SL(2,\R)$ $ \ti SL(2,\R)$ relations for the
affine-Virasoro
connections. For example, the right index or $L$-relations,
$$
A_g \sum_{i=1}^n W_{j_1  \ldots j_q, i_1 \ldots i_p i } = 0
\eqno(4.9a)$$
$$
A_g [ \sum_{i=1}^n z_i W_{j_1 \ldots j_q, i_1 \ldots i_p i} +
(  \sum_{i=1}^n \D_i + p ) W_{j_1 \ldots j_q ,i_1 \ldots i_p} ] = 0
\eqno(4.9b)$$
$$
\eqalign{
\;\;\;A_g [ \sum_{i=1}^n z_i^2 W_{j_1 \ldots j_q,i_1 \ldots i_p i}
& + 2( \sum_{i=1}^n z_i \D_i + \sum_{\n =1}^p z_{i_{\n}} )
W_{j_1 \ldots j_q,i_1 \ldots i_p } \cr
& + 2 \sum_{\n =1}^ p (\D_{i_{\n}} + \sum_{\m = \n +1}^p \d_{i_{\n},i_{\m}} )
W_{j_1 \ldots j_q, i_1 .. \hat{i}_{\n} .. i_p } ] = 0 \cr}
 \eqno(4.9c)$$
are obtained from (4.8b) using $\pa_i A = A W_i$, eq.(2.6) and
the Ward identities (3.17a). In (4.9c), the omission of an
index is denoted by a hat. Analogous left index or $\tL$-relations with
$\D \ra \tD$  are obtained from eq.(4.8a) and $\bpa_i A =A \bW_i $.

In the remainder of this section, we use the results,
$$
\bW_i(\bz,z) = \Fb^{-1}(\bz,z)  \pa_i \Fb (\bz,z)\;\;\;\;,\;\;\;\;
W_i (\bz,z) = F^{-1} (\bz,z) \pa_i F (\bz,z)
\eqno(4.10a)$$
$$
Q_a (\bz,z) = F^{-1} (\bz,z) Q^g_a F(\bz,z)  = \Fb^{-1} (\bz,z)Q^g_a \Fb
(\bz,z)
\eqno(4.10b)$$
$$
\Fb(\bz,z) = \sum_{q=0}^{\infty} {1 \over q !} \sum_{j_1 \ldots
 j_q}
\prod_{\n=1}^q (\bz_{j_\n} - z_{j_\n } ) W_{j_1 \ldots j_q,0} (z)
\eqno(4.10c)$$
$$ F(\bz,z) = \sum_{p=0}^{\infty} {1 \over p !} \sum_{i_i \ldots i
_p}
\prod_{\m=1}^p (z_{i_\m} - \bz_{i_\m } ) W_{0,i_1 \ldots i_p} (\bz)
\eqno(4.10d)$$
to translate some of what is known about the affine-Virasoro
connections into the language
of flat connections.

\ni 1. Pinched flat connections. Using (4.10) or (2.7), we
obtain the pinched flat
connections,
$$
\bW_i (z,z) = W_{i,0} (z) = 2 \tL^{ab} \sum_{j \neq i} {\T_a^i \T_b^j \over
z_{ij}}
\eqno(4.11a)$$
$$
W_i (z,z) = W_{0,i} (z) = 2 L^{ab} \sum_{j \neq i} {\T_a^i \T_b^j \over
z_{ij}}
\eqno(4.11b)$$
$$
\bW_i (z,z) + W_i (z,z) = W_i^g (z)
 \eqno(4.11c)$$
from the known form \cite{WI} of the first-order affine-Virasoro connections.

\ni 2. $\tL$ and $L$ dependence \cite{WI}.
The one-sided  connections $W_{j_1 \ldots j_q,0 }(\tL,z)$,
and $W_{0,i_1 \ldots i_p} (L,z)$ are functions of $\tL$ and $L$
respectively, so the evolution operators and flat connections are also
functions only of $\tL$ or $L$,
$$
\Fb (\tL ,\bz,z) \;\;\;\;, \;\;\;\; F(L,\bz,z)
\eqno(4.12a)$$
$$
\;\;\;\; \bW_i (\tL,\bz,z) \;\;\;\;,\;\;\;\; W_i (L,\bz,z) \;\;\;\;.
\eqno(4.12b)$$

\ni 3. K-conjugation covariance. Under K-conjugation, the affine-Virasoro
connections
satisfy \cite{WI}
$$
 W_{j_1 \ldots j_q,0} (\tL=L_*) = W_{0, j_1 \ldots j_q} (L=L_*)
\;\;,\;\;
 W_{0,i_1 \ldots i_p} (L=L_*) = W_{i_1 \ldots i_p,0} (\tL=L_*)
\eqno(4.13) $$
so the evolution operators and flat connections satisfy,
$$
\Fb (\tL =L_*,\bz,z) = F(L=L_* ,z,\bz)
\;\;\;\;,\;\;\;\;
F (L =L_*,\bz,z) = \Fb (\tL=L_* ,z,\bz)
\eqno(4.14a)$$
 $$
\bW_i (\tL= L_* ,\bz,z) = W_i (L = L_*,z,\bz)
\;\;,\;\;
W_i (L=L_*,\bz,z) = \bW_i (\tL = L_*,z,\bz) \;\;.
\eqno(4.14b)$$
 From these relations, we also learn that the biconformal correlators and
non-local conserved quantities satisfy
$$
A(\bz,z) \ve_{\tL \lra L}  = A(z,\bz)
\eqno(4.15a)$$
$$
Q_a(\bz,z) \ve_{\tL \lra L}  = Q_a(z,\bz)   \;\;\;\;, \;\; a= 1,\ldots,
{\rm dim}\,g
\eqno(4.15b)$$
under exchange of the K-conjugate theories.

\ni 4. Crossing symmetry. The affine-Virasoro connections satisfy the
crossing symmetry \cite{WI2}
$$
W_{j_1 \ldots j_q, i_1 \ldots i_p} (z) \ve_{k \lra l} =
W_{j_1 \ldots j_q, i_1 \ldots i_p} (z)
\eqno(4.16a)$$
$$
\T^ k \lra \T^l \;\;\;\;,\;\;\;\;  z_k \lra z_l \;\;\;\;,\;\;\;\;
i \lra l\;{\rm or}\; k\;\, {\rm when}\;\,i=k\;{\rm or}\;l
\eqno(4.16b)$$
where $k\lra l$ means exchange of $\T$'s, $z$'s
 and indices, as given in (4.16b).
It follows from (4.10) that the
evolution operators, flat connections and non-local conserved quantities
 satisfy the crossing relations
$$
\Fb (\bz,z) \ve_{k\lra l} = \Fb(\bz,z) \;\;\;\;, \;\;\;\;
F (\bz,z) \ve_{k\lra l} = F(\bz,z)
\eqno(4.17a)$$
$$
\bW_i (\bz,z) \ve_{k\lra l} = \bW_i (\bz,z) \;\;\;\;, \;\;\;\;
W_i  (\bz,z) \ve_{k\lra l} = W_i (\bz,z)
\eqno(4.17b)$$
$$
Q_a(\bz,z) \ve_{k \lra l} = Q_a (\bz,z)
\eqno(4.17c)$$
$$
\T^ k \lra \T^l \;\;\;\;,\;\;\;\;  \bz_k \lra \bz_l \;\;\;\;,\;\;\;\;
z_k \lra z_l \;\;\;\;,\;\;\;\;
i \lra l\;{\rm or}\; k\;\, {\rm when}\;\,i=k\;{\rm or}\;l
\eqno(4.17d)$$
where $k \lra l$ now also includes the exchange of $\bz$'s, as given in
(4.17d).

\ni 5. High-level flat connections. On simple $g$, we know the high-level
affine-Virasoro connections \cite{WI2}
$$
 W_{j_1 \ldots j_q,0} = \left( \prod_{r=1}^{q-1} \pa_{j_r} \right) W_{j_q,0} +
\cO (k^{-2}) \;\;\;,\;\; q \geq 1
\eqno(4.18a)$$
$$
 W_{0,i_1 \ldots i_p} = \left( \prod_{r=1}^{p-1} \pa_{i_r} \right) W_{0,i_p} +
\cO (k^{-2}) \;\;\;,\;\; p \geq 1
\eqno(4.18b)$$
$$
 W_{j_1 \ldots j_q,i_1 \ldots i_p} =\cO (k^{-2})\;\;\;,\;\; q,p \geq 1
\eqno(4.18c)$$
$$
\tL^{ab} = { \tilde{P}^{ab} \over 2k} + \cO (k^{-2}) \;\;\;\;,\;\;\;\;
L^{ab} = { P^{ab} \over 2k} + \cO (k^{-2})
\eqno(4.18d)$$
where $W_{i,0}$ and $W_{0,i}$ are given in (4.11a,b)
and $\tilde{P}$ and $P$ are
the high-level projectors [17,18]
of the $\tL$ and the $L$ theory respectively. This
translates to the high-level form of the evolution operators,
$$
 \Fb(\bz,z) = \one + { \tilde{P}^{ab} \over k} \sum_{i <j}
  \T_a^i \T_b^j
\ln \left( {\bz_{ij} \over z_{ij}} \right)
+ \cO (k^{-2} )
\eqno(4.19a)$$
$$
 F(\bz,z) = \one + {P^{ab} \over k}
 \sum_{i < j}
  \T_a^i \T_b^j
\ln \left( {z_{ij} \over \bz_{ij}} \right)
+ \cO (k^{-2} )
\eqno(4.19b)$$
and so we obtain
$$
\bW_i(\bz,z) = W_{i,0} (\bz) + \cO (k^{-2})
= {\tilde{P}^{ab} \over k} \sum_{j \neq i} {\T_a^i \T_b^j \over z_{ij}} +
\cO (k^{-2} )
\eqno(4.20a)$$
$$
W_i(\bz,z) = W_{0,i} (z) + \cO (k^{-2})
= {P^{ab} \over k} \sum_{j \neq i} {\T_a^i \T_b^j \over z_{ij}} + \cO (k^{-2})
\eqno(4.20b)$$
for the high-level flat connections of ICFT.

Using (4.10b-d)
and (4.19), we
also obtain the
high-level forms of the non-local conserved quantities
$$
Q_a(\bz,z) = Q^g_a + [Q^g_a, {P^{ab} \over k} \sum_{i <j}
\ln \left( { z_{ij} \over \bz_{ij} } \right) \T_a^i \T_b^j ]
+ \cO (k^{-2} )
\eqno(4.21a)$$
$$
 \;\; =Q^g_a + [Q^g_a, {\tilde{P}^{ab} \over k} \sum_{i <j}
\ln \left( { \bz_{ij} \over z_{ij} } \right) \T_a^i \T_b^j ]
+ \cO (k^{-2} )
\eqno(4.21b)$$
where the equality of the two forms is easily verified with
$\tL + L = L_g$.

In Section 7, we use these results to solve the system of flat connections
for the high-level $n$-point correlators of ICFT.

\section{Flat Connections for Cosets and Nests}

The affine-Sugawara nests [9-11,13,14] generalize the $g/h$ coset
constructions [2,5,8] to CFT's on the subgroup nests\footnote{The higher nests
($n \geq 2$) are tensor product CFT's \cite{WI2}, but this will not be needed
here.}
$g \supset h_1 \supset \ldots \supset h_n $.
The all-order affine-Virasoro connections
of the cosets and nests were computed
in Ref.\cite{WI2}, employing an iterative scheme based on  K-conjugation and
the solution of the consistency relations
(2.10).
Using these results, we may obtain the flat connections of the cosets
and nests from eq.(4.10).

Instead, we develop here a related iterative method which works directly
with the flat connections. In  particular, the scheme uses only the relations,
$$
\bW_i [\tL= L_*,\bz,z] = W_i [L=L_*,z,\bz]
\eqno(5.1a)$$
$$
\pa_i F = F W_i \;\;\;\;,\;\;\;\; F(z,z) = \one
\eqno(5.1b)$$
$$
\bW_i (\bz,z) = F^{-1} (\bz,z) (\bpa_i + W_i^g(\bz))F(\bz,z)
\eqno(5.1c)$$
collected from above, where (5.1a) is the K-conjugation covariance of
the flat connections.

\ni \un{$\tL = L_g,\;L=0$} \nl
For the trivial theory $L=0$, we begin with the trivial connection,
$$
W_i [L=0,\bz,z]=0 \;\;\;\;.
\eqno(5.2)$$
Then we may compute
$$
F[L=0,\bz,z] = \one
\eqno(5.3a)$$
$$
\bW_i[\tL=L_g,\bz,z] = W_i^g (\bz)
\eqno(5.3b)$$
from (5.1b) and (5.1c), where $W_i^g (z)$ is the KZ connection in
(4.3a).

\ni \un{$\tL=L_{g/h}$, $L=L_h$} \nl
When $h \subset g$, we may rename the groups to obtain
$$  W_i [ L=L_h,\bz,z] = W_i^h (z)
= 2 L^{ab}_h \sum_{j \neq i} { \T_a^i \T_b^j \over z_{ij} }
\eqno(5.4)$$
from (5.3b) and (5.1a). Then, we compute from (5.1b) and
(5.1c) that
$$
F [ L=L_h,\bz,z] =  A_h (z,\bz)
= A_h^{-1} (\bz,z)
\eqno(5.5a)$$
$$
\eqalign{
  \bW_i  [\tL = L_{g/h},\bz,z] & = A_h(\bz,z) (\bpa_i + W_i^g(\bz) ) A_h(z,\bz)
\cr
& =
  A_h (\bz,z) W_{i}^{g/h}  (\bz) A_h (z,\bz) \cr}
\eqno(5.5b)$$
where $A_h$ is the evolution operator (4.5) on $h$
and $W_i^{g/h} = W_i^g - W_i^h$.

\ni \un{$\tL=L_{g/h_1/h_2}, \;L=L_{h_1/h_2}$} \nl
When $g \supset h_1 \supset h_2$, we obtain
 $$ W_i [ L=L_{h_1/h_2},\bz,z]
  = A_{h_2} (z,\bz) W_{i}^{h_1/h_2}  (z) A_{h_2} (\bz,z)
\eqno(5.6a)$$
$$
F [ L=L_{h_1/h_2},\bz,z] = A_{h_1}(z,\bz) A_{h_2} (\bz,z)
\eqno(5.6b)$$
$$
  \bW_i  [ \tL=L_{g/h_1/h_2},\bz,z]
  = A_{h_2} (z,\bz) A_{h_1} (\bz,z) W_{i}^{g/h_1}  (\bz) A_{h_1} (z,\bz)
A_{h_2}(\bz,z) + W_i^{h_2} (\bz)
\eqno(5.6c)$$
where (5.6a) follows from (5.5b) and (5.1a),
and (5.6b,c) are computed from (5.1b) and (5.1c)
respectively.

 From these examples, it is clear that we are following an iterative procedure
whose general form is
$$
\bW_i [ \tL = L_{g/h_1/\ldots /h_n} ] \ra
\bW_i [\tL = L_{h_1 /\ldots /h_{n+1}} ]
$$
$$
\ra W_i [ L= L_{h_1 / \ldots / h_{n+1}} ] \ra F [ L =L_{h_1 / \ldots /h_{n+1}}]
$$
$$
\ra \bW_i [ \tL = L_{g /h_1 /\ldots /h_{n+1}} ] \;\;\;\;.
\eqno(5.7)$$
The first step in (5.7) is a renaming of the groups, followed by the
application of eqs.(5.1a,b,c) in that order.

Continuing this iteration, we find
 that the evolution operators of the general nest have the form
$$
\eqalignno{F[L=L_{h_1/\ldots/h_{2n+1}},\bz,z]
 &= A_{h_1}(z,\bz) A_{h_2} (\bz,z) \cdots A_{h_{2n+1}} (z,\bz)
&(5.8a) \cr
F[L=L_{h_1/\ldots/h_{2n}},\bz,z]
 & = A_{h_1}(z,\bz) A_{h_2} (\bz,z) \cdots A_{h_{2n}} (\bz,z)
&(5.8b) \cr
\Fb [\tL=L_{g/h_1/\ldots /h_n } ,\bz,z]& =A_g(\bz,z)
F[ L=L_{h_1/\ldots /h_n} ,\bz,z]
&(5.8c)\cr} $$
where $A_{h_i}(z,z_0)$ is the evolution operator on
$h_i$,
defined as shown in eq.(4.5). It follows
from (5.8) and (5.1c)
that the flat connections satisfy the recursion relations,
$$
\eqalign{
 \bW_i  & [\tL=L_{g/h_1/\ldots /h_{2n+1} },\bz,z] \cr
& = A_{h_{2n+1}} (\bz,z) \left( \bW_i (\bz,z) [\tL= L_{g/h_1/\ldots /h_{2n}}]
- W_i^{h_{2n+1}} (\bz) \right) A_{h_{2n+1}} (z,\bz) \cr}
\eqno(5.9a)$$
$$
\eqalign{
 \bW_i  & [\tL=L_{g/h_1/\ldots /h_{2n} },\bz,z] \cr
& = A_{h_{2n}} (z,\bz)  \bW_i (\bz,z) [\tL= L_{g/h_1/\ldots /h_{2n-1}}]
A_{h_{2n}}(\bz,z) + W_i^{h_{2n}} (\bz)  \cr}
\eqno(5.9b)$$
which generate the flat nest connections to any desired depth, starting
from eq.(5.3b).

We remark that these connections can also be obtained in an alternate iteration
scheme based on the relations
$$
\bW_i[\tL  = L_*,\bz,z] = W_i[L=L_*,z,\bz]
\eqno(5.10a)$$
$$
(\pa_i + W_i) \bW_j =( \bpa_j + \bW_j ) W_i
\eqno(5.10b)$$
$$
\bW_i (\tL,z,z) = W_{i,0} (\tL,z) \;\;\;\;,\;\;\;\;
W_i (L,z,z) = W_{0,i} (L,z) \;\;\;\;.
\eqno(5.10c)$$
In this scheme, one solves the cross-flatness condition (5.10b)
for $\bW_i$ at
each depth of $g/h_1/\ldots /h_n$, fixing the constants of
integration by the boundary conditions (5.10c).

As a check on these results, we compute the biconformal nest correlators, using
eqs.(2.19), (4.5) and
the evolution operators in (5.8). We obtain
$$
\eqalignno{
A &[ \tL=L_{g/h_1/\ldots /h_{2n+1}},L=L_{h_1/\ldots /h_{2n+1}}]
&(5.11a) \cr
&= A_g (\bz) A_{h_1}(z,\bz) A_{h_2} (\bz,z) \cdots A_{h_{2n+1}}(z,\bz)
&(5.11b) \cr
& =A_g(z_0) [ A_{g/h_1}(\bz,z_0) A_{h_1/h_2}(z,z_0) \cdots A_{h_{2n}/h_{2n+1}}
(\bz,z_0) A_{h_{2n+1}}(z,z_0) ] \;\;\;\;\;\;
&(5.11c)\cr } $$
$$
\eqalignno{ A & [ \tL=L_{g/h_1/\ldots /h_{2n}},L=L_{h_1/\ldots /h_{2n}}]
&(5.11d) \cr
&= A_g (\bz) A_{h_1}(z,\bz) A_{h_2} (\bz,z) \cdots A_{h_{2n}}(\bz,z)
&(5.11e) \cr
&=A_g(z_0) [ A_{g/h_1}(\bz,z_0) A_{h_1/h_2}(z,z_0) \cdots A_{h_{2n-1}/h_{2n}}
(z,z_0) A_{h_{2n}}(\bz,z_0) ] \;\;\;\;\;\;
&(5.11f) \cr} $$
where the definition
$$
A_{h_i/h_{i+1}}(z,z_0) \equiv A_{h_i}(z,z_0) A_{h_{i+1}}^{-1}(z,z_0)
\eqno(5.12)$$
is used in (5.11c) and (5.11f). The results in (5.11c,f)
are precisely those obtained
in Ref.\cite{WI2}, while the alternate
 forms in (5.11b,e) show explicitly that
the biconformal correlators are independent of the reference point $z_0$
\cite{WI2}.

We turn now to the non-local conserved quantities $Q_a=F^{-1}Q_a^g F$ of the
cosets
and nests.
 In the form (5.11c,f), the biconformal
nest correlators satisfy the non-local
conservation laws
$$
A(\bz,z) Q_a(\bz,z) =0 \;\;\;\;,\;\;a =1 , \ldots , {\rm dim}\,g
\eqno(5.13)$$
so long as the affine-Sugawara correlator $A_g^{\a}(z_0)$ satisfies the global
Ward identity $A_g(z_0) \sum_{i=1}^n \T_a^i =0 $. In further detail, if one
considers only the differential system (3.16a) of flat connections, the
solutions for the biconformal nest correlators
are those in (5.11c,f), with $A_g(z_0)$
replaced by an arbitrary initial condition $A(z_0)$. The initial condition is
then fixed to $A_g(z_0)$
by requiring the non-local conservation law (5.13).

We also remark that the non-local generators $Q_a ,\; a \in h$
simplify to the usual global generators of $h$ when $h \subset g$ is an
ordinary symmetry of the construction. To see this, note first that
$$
[ Q_a^g , W_j^{h_i} (z) ] =[ Q_a^g, A_{h_i} (\bz,z) ] = 0 \;\;\;\;, \;\;
a \in h_k \;\;\;,\;\; k \geq i
\eqno(5.14a)$$
$$
[ Q_a^g , W_j^{h_i} (z) ] = [Q_a^g , A_{h_i} (\bz,z) ] = 0 \;\;\;\;,\;\;
a \in h_n \;\;\;\;.
\eqno(5.14b)$$
$$
g \supset h_1 \supset \ldots \supset h_n \;\;\;\;, \;\;\;\; g \equiv h_0
\eqno(5.14c)$$
where (5.14a) follows from the embedding order of the subgroups and
(5.14b) is included in (5.14a). It follows from (5.8),
(4.10b) and
(5.14b) that
$$
\tL = L_{g /h_1 /\ldots /h_n} \;\;\;\;,\;\;\;\;
L = L_{h_1 / \ldots /h_n}
\eqno(5.15a)$$
$$
Q_a (\bz,z) = Q_a^g \;\;\;\;,\;\;a \in h_n
\eqno(5.15b)$$
while the other generators $Q_a(\bz,z)
,\; a \in g/h_n$ remain generically non-local.
The coordinate-independent form of the $h$ generators in
(5.15b) is in accord with the known Lie $h$-invariance
of the cosets and nests, and should persist for all Lie $h$-invariant conformal
field theories \cite{Lieh}.

\section{Induction to Coset CFT's}

The coset correlators \cite{WI}, which describe the coset CFT's, are obtained
by
factorization [13,14] from the biconformal correlators of $h$ and the $g/h$
coset
constructions. In this section, we note that the flat connections and non-local
conserved quantities of the biconformal correlators induce related flat
connections and non-local conserved quantities for the
coset CFT's themselves.

To warm up, we consider first the case of the affine-Sugawara construction
on $g$,
which is K-conjugate to the trivial theory. In this case, the results
of Section 5 reduce to
$$
\tL = L_g \;\;\;\;, \;\;\;\; L=0
\eqno(6.1a)$$
$$
\bW_i (\bz,z) = W_i^g (\bz) \;\;\;\;,\;\;\;\; W_i(\bz,z) =0
\eqno(6.1b)$$
$$
\bpa_i A(\bz,z) = A(\bz,z) W_i^g (\bz) \;\;\;\;, \;\;\;\;
\pa_i A(\bz,z) = 0
\eqno(6.1c)$$
$$
A(\bz,z) = A_g (\bz)
\eqno(6.1d)$$
$$
 Q_a (\bz,z) = Q^g_a = \sum_{i=1}^n\T_a^i \;\;\;\;,\;\; a = 1 ,\ldots ,
{\rm dim}\, g
\eqno(6.1e)$$
so that the biconformal correlator is the affine-Sugawara correlator on
$g$ and $A (\bz,z)\,Q_a(\bz,z)=0$
is the usual $g$-global Ward identity.
At the risk of belaboring the simplest case, this example shows clearly that
the KZ equations are included in the general system
(3.16) of flat connections, and
that the general system is a large set of generalized KZ equations.

For  $h$ and the $g/h$ coset constructions, the results of Section 5 reduce
to
$$
\tL= L_{g/h} \;\;\;\;,\;\;\;\; L=L_h
\eqno(6.2a)$$
$$
\bW_i (\bz,z) = A_h (\bz,z) W_i^{g/h} (\bz) A_h (z,\bz) \;\;\;\;,\;\;\;\;
W_i(\bz,z) = W_i^h (z)
\eqno(6.2b)$$
$$
A^{\a}(\bz,z) = A_{g/h}^{\b}(\bz,z_0) A_h (z,z_0)_{\b}{}^{\a}
\eqno(6.2c)$$
$$
A^{\a}_{g/h} (\bz,z_0) \equiv A_g^{\b} (z_0) A_{g/h} (\bz,z_0)_{\b}{}^{\a}
\eqno(6.2d)$$
$$
Q_a (\bz,z) = \left\{
\matrix{ Q^g_a & ,\;\;a \in h \cr
A_h(\bz,z) Q^g_a A_h(z,\bz) & ,\;\;a \in g/h \cr}
\right.
\eqno(6.2e)$$
where $A_{g/h}^{\a}$ in (6.2d) are the coset
correlators \cite{WI}, whose corresponding coset
blocks [19,13,20] were first discussed by Douglas.

In this case, we remark first that the coset correlators are known to satisfy
the
coset equations \cite{WI},
$$
\bpa_i A_{g/h}^{\a}(\bz,z_0)  = A_{g/h}^{\b} (\bz,z_0)
W_i [g/h, \bz,z_0 ]_{\b}{}^{\a}
\eqno(6.3a)$$
$$
W_i[g/h,\bz,z_0] = A_h (\bz,z_0) W_i^{g/h} (\bz) A_h^{-1} (\bz,z_0)
\eqno(6.3b)$$
where the dressed coset connections $W_i[g/h]$ in (6.3b)
are flat connections. Comparing the flat connections in (6.2b) and
(6.3b), we see that the dressed coset connections are induced from
the barred flat connections
by choosing $z$ to be the regular reference point $z_0$,
$$
W_i[g/h,\bz,z_0] = \bW_i [\tL = L_{g/h}, \bz,z_0]   \;\;\;\; .
 \eqno(6.4)$$
See Section 8 for further remarks on induced flat connections for the
correlators of all ICFT.

Our second remark is that the coset correlators also enjoy a set of non-local
conserved generators of $g$, induced from the non-local conserved generators
of the biconformal correlators. To see the induction, we need the identity
$$
A_h (z,z_0) Q_a (\bz,z) = Q_a(\bz,z_0) A_h(z, z_0)
\eqno(6.5)$$
which follows from the properties (4.5b,c)
of $A_h$. Then, beginning with the
non-local identity for the biconformal correlator, we have
$$
\eqalign{
0 & = A_{g/h}^{\g} (\bz,z_0) A_h (z,z_0)_{\g}{}^{\b} Q_a (\bz,z)_{\b}{}^{\a}
\cr
& = A_{g/h}^{\g}(\bz,z_0) Q_a(\bz,z_0)_{\g}{}^{\b}
 A_h(z,z_0)_{\b}{}^{\a} \;\;\;\;. \cr}
 \eqno(6.6)$$
This tells us that the induced generators $Q_a(\bz,z_0)$
are non-local conserved generators of $g$ for the
coset correlators,

$$
A_{g/h}^{\b} (\bz,z_0) Q_a (\bz,z_0)_{\b}{}^{\a} = 0
\eqno(6.7a)$$
$$
[ Q_a(\bz,z_0) , Q_b(\bz,z_0) ] = i f_{ab} {}^c Q_c(\bz,z_0)
\eqno(6.7b)$$
$$
Q_a(\bz,z_0) = \left\{
\matrix{ Q^g_a & ,\;\;a \in h \cr
A_h(\bz,z_0) Q^g_a A_h(z_0,\bz) & ,\;\; a \in g/h \cr}
\right.
\eqno(6.7c)
$$
because $A_h$ is invertible. The result (6.7)
can also be verified directly from the form of the coset correlator in
(6.2d).

As anticipated, eq.(6.7) shows the well-known global
$h$-invariance of the coset correlators \cite{WI}, supplemented now with the
new
non-local conserved generators of $g/h$.

\section{The High-Level Biconformal Correlators of ICFT}

Collecting the high-level results from Section 4, we wish to solve the
high-level system of flat  connections,
$$
\bpa_i  A (\bz,z) = A(\bz,z)  \left( {\tilde{P}^{ab} \over k} \sum_{j \neq i}
{ \T_a^i \T_b^j \over \bz_{ij} }
+ \cO (k^{-2} ) \right )
\eqno(7.1a)$$
$$
\pa_i  A (\bz,z) = A (\bz,z)\left( {P^{ab} \over k} \sum_{j \neq i}
{ \T_a^i \T_b^j \over z_{ij} }
+ \cO (k^{-2} ) \right )
\eqno(7.1b)$$
$$
A (\bz,z) \, Q_a (\bz,z) = \cO (k^{-2})
\eqno(7.1c)$$
$$
\eqalignno{Q_a(\bz,z) &= Q^g_a + [Q^g_a, {P^{ab} \over k} \sum_{i <j}
\T_a^i \T_b^j  \ln \left( { z_{ij} \over \bz_{ij} } \right)
 ]
+ \cO (k^{-2} )
&(7.1d) \cr
 &  =Q^g_a + [Q^g_a, {\tilde{P}^{ab} \over k} \sum_{i <j}
\T_a^i \T_b^j  \ln \left( { \bz_{ij} \over z_{ij} } \right)
 ]
+ \cO (k^{-2} )
&(7.1e) \cr} $$
where $\tilde{P}$ and $P$ are the high-level projectors of the $\tL$ and the
$L$
theory respectively. The solutions of this system are the high-level $n$-point
biconformal
correlators of ICFT,
$$
A(\bz,z) = A_g(z_0)
\left( \one + \sum_{i<j} \T_a^i \T_b^j \left[
{ P^{ab} \over k}  \ln \left( {z_{ij} \over z_{ij}^0} \right)
+ {\tilde{P}^{ab} \over k} \ln \left( {\bz_{ij} \over z_{ij}^0
} \right)  \right]   + \cO (k^{-2}) \right)
\eqno(7.2a)$$
$$
A(z_0,z_0) = A_g(z_0) \;\;\;\;,\;\;\;\; A_g(z_0) Q^g_a =0
\eqno(7.2b)$$
where $z_0$ is a regular reference point.

In further detail, the solution to the differential system (7.1a,b)
gives (7.2a) with
the left factor as  an undetermined row vector
$A^{\b} (z_0) $, instead of $A_g^{\b}(z_0)$.
 The row vector is then fixed to be the affine-Sugawara correlator
$ A_g(z_0)$ by the non-local conservation law (7.1c).

Here is a partial list of the properties of the high-level biconformal
correlators in (7.2).

\ni A. {\it An alternate derivation}. The same result
(7.2)
is obtained by summing the partially-factorized form of the biconformal
correlators \cite{WI2}
$$
\eqalign{ A^{\a} & (\bz,z) = \cr
 & \sum_{q,p=0}^{\infty} \frac{1}{q!}
\sum_{j_1 \ldots j_q}^n \frac{1}{p!}
\sum_{i_1 \ldots i_p}^n \prod_{\m=1}^q (\bz_{j_{\m}} - z^0_{j_{\m}}) \,
[ A_g^{\b}(z_0) W_{j_1 \ldots j_q,i_1 \ldots i_p}(z_0 {)_{\b}}^{\a}]
 \prod_{\n=1}^p (z_{i_{\n}} -z^0_{i_{\n}} ) \cr}
\eqno(7.3)$$
using the high-level form of the affine-Virasoro
 connections in (4.18).

\ni B. $z_0$-{\it independence}. The high-level biconformal correlators
(7.2) are independent of the reference point $z_0$,
$$
{\pa \over \pa z_i^0} A(\bz,z) = \cO (k^{-2} )
\eqno(7.4)$$
as they should be. To see this, we may use the high-level form of the
KZ equation
$$
{\pa \over \pa z_i^0} A_g(z_0) = A_g(z_0) \left( {(\tilde{P} + P )^{ab}
\over k} \sum_{j \neq i} {\T_a^i \T_b^j \over z^0_{ij} } + \cO (k^{-2}) \right)
\eqno(7.5)$$
or we may rearrange the result in the equivalent forms,
$$
\eqalign{
A(\bz.z) & = A_g(z) \left[ \one + {\tilde{P}^{ab}  \over k} \sum_{i<j}
\T_a^i \T_b^j
\ln
\left( { \bz_{ij} \over z_{ij} } \right)
\right] + \cO(k^{-2} )
\cr
 & = A_g(\bz) \left[ \one + {P^{ab}  \over k} \sum_{i<j}
\T_a^i \T_b^j
\ln
\left( { z_{ij} \over \bz_{ij} } \right)
\right] + \cO(k^{-2} )
\cr }
\eqno(7.6)$$
which also follow from eqs.(2.19) and (4.19).

\ni C. $SL(2,\R) \ti SL(2,\R)$ {\it covariance}. Using the global Ward identity
(7.2b) and the
conformal weights of the broken
affine primary fields \cite{WI}
$$
( \tL^{ab} \T_a \T_b )_{\a}{}^{\b} =\d_{\a}^{\b} \tD_{\a} \;\;\;\;,\;\;\;\;
( L^{ab} \T_a \T_b )_{\a}{}^{\b} = \d_{\a}^{\b} \D_{\a}
\eqno(7.7)$$
we verify that the biconformal correlators satisfy the
$SL(2,\R) \ti SL(2,\R)$ relations,
$$
\eqalignno{\sum_{i=1}^n \bpa_i A^{\a}
= \sum_{i=1}^n (\bz_i \bpa_i + \tD_{\a_i} ) A^{\a}
& = \sum_{i=1}^n (\bz_i^2 \bpa_i + 2 \bz_i \tD_{\a_i} ) A^{\a}  =\cO (k^{-2})
&(7.8a) \cr
 \sum_{i=1}^n \pa_i A^{\a}
= \sum_{i=1}^n (z_i \pa_i + \D_{\a_i} ) A^{\a}
& = \sum_{i=1}^n (z_i^2 \pa_i + 2 z_i \D_{\a_i} ) A^{\a}  =\cO (k^{-2})
&(7.8b) \cr }$$
in accord with eq.(4.8).

\ni D. {\it Factorized form}. The biconformal
correlators are easily written in the
factorized form
$$
\eqalign{ A(\bz,z) = A_g(z_0)
 & \left[ \one  +
{ \tilde{P}^{ab} \over k}  \sum_{i <j} \T_a^i \T_b^j
\ln \left( {\bz_{ij} \over z_{ij}^0} \right)
\right]
\cr
\ti & \left[ \one +
{P^{ab} \over k} \sum_{i <j} \T_a^i \T_b^j
\ln \left(
{z_{ij} \over z_{ij}^0
} \right)  \right]  + \cO (k^{-2} )   \cr }
\eqno(7.9)$$
since the correction terms are $\cO( k^{-2})$.
Using this result, we may obtain the high-level conformal correlators
$\bA (\bz,z_0) $ from the factorization \cite{WI}
$$
A^{\a} (\bz,z) = \bA^{\b} (\bz,z_0) A(z,z_0)_{\b}{}^{\a} + \cO (k^{-2} )
\;\;\;\;.
\eqno(7.10)$$
These correlators are given and discussed in the following section.

\section{The High-Level Conformal Correlators of ICFT}

Comparing the factorized forms (7.9) and (7.10) of the biconformal correlators,
we read off
the high-level $n$-point conformal correlators of ICFT,
$$
\bA (\bz,z_0) =
 A_g(z_0)
 \left[ \one +
{ \tilde{P}^{ab} \over k}
 \sum_{i <j} \T_a^i \T_b^j \ln \left( {\bz_{ij} \over z_{ij}^0} \right)
\right] + \cO (k^{-2} )
\eqno(8.1a)$$
$$
\tL^{ab} = { \tilde{P}^{ab} \over 2k } + \cO (k^{-2} )
\;\;\;\;. \eqno(8.1b)$$
Here is a partial list of properties of these correlators.

\ni A. {\it High-level cosets and nests}.
Using the high-level forms of the evolution operators of $g$ and $h$,
$$
A_g(z,z_0) = \one + {P_g^{ab} \over k} \sum_{i < j} \T_a^i \T_b^j \ln \left(
z_{ij} \over z_{ij}^0 \right) + \cO (k^{-2})
\eqno(8.2a)$$
$$
A_h^{-1}(z,z_0) = \one - {P_h^{ab} \over k} \sum_{i <j} \T_a^i \T_b^j \ln
\left(
z_{ij} \over z_{ij}^0 \right) + \cO (k^{-2})
\eqno(8.2b)$$
we verify that the
high-level forms of the coset correlators (6.2d) are
correctly included in (8.1) when
$\tilde{P}=P_{g/h}= P_g -P_h$.
Similarly, the high-level form of the nest correlators are obtained from
(8.1) when
$$
\tilde{P} = P_{g/ h_1 /\ldots /h_n} = P_g + \sum_{j=1}^n (-)^j P_{h_j}
\eqno(8.3)$$
and we have checked that the four-point conformal blocks of these correlators
agree
with the nest conformal blocks obtained in Ref.\cite{WI2}.

\ni B. {\it Induced flat connections for ICFT}. It was suggested in
Ref.\cite{WI}
that the $n$-point correlators of ICFT might satisfy a PDE with flat
connections,
$$
\bpa_i \bA = \bA W_i[\tL] \;\;\;\;,\;\;\;\;
W_i[\tL] = { \tilde{P}^{ab} \over k}
\sum_{j \neq i} {\T_a^i \T_b^j \over \bz_{ij} } + \cO (k^{-2})
\eqno(8.4)$$
and our result (8.1) was obtained in eq.(14.3) of \cite{WI} as the solution to
this equation. The derivation of (8.1) in the present paper confirms this
conjecture to the indicated order, and we observe that the conjectured form
of the flat connection in (8.4)  is precisely the high-level form
of the flat connection $\bW_i (\bz,z_0)$.

Considered with the result (6.4) for the induced flat connections of
the finite-level coset correlators, this observation suggests a stronger form
of the conjecture, namely that
$$
\bW_i (\tL, \bz,z_0) \;\;\;\;,\;\;\;\; W_i(L,z_0,z)
\eqno(8.5)$$
are the finite-level
induced flat connections of the $\tL$ and the $L$ theory respectively, where
$z_0$ is a regular reference point.
This conjecture should be investigated vis-a-vis finite-level
factorization, as discussed in Ref.\cite{WI2}.

\ni C. $SL(2,\R)$ {\it covariance}. With Ref.\cite{WI}, we verify the
expected $SL(2,\R)$ relations,
$$
\sum_{i=1}^n \bpa_i \bA (\bz,z_0) = \sum_{i=1}^n (\bz_i \bpa_i + \tD_{\a_i})
\bA (\bz,z_0) = \sum_{i=1}^n (\bz_i^2 \bpa_i + 2 \bz_i \tD_{\a_i} )
\bA (\bz,z_0) = \cO (k^{-2} )
\eqno(8.6)$$
using the global Ward identity (7.2b) and the known conformal weights
$\tD = {\rm diag}(\tL^{ab} \T_a \T_b )$ of the broken affine primary fields.

\ni D. {\it Two- and three-point correlators}.
Choosing $n=2$ in eq.(8.1), we obtain the high-level two-point correlators
of ICFT,
$$
\bA^{\a_1 \a_2} (\bz_1\,z_1^0\,\T^1,\bz_2\, z_1^0 \,\T^2) =
{\eta^{\a_1 \a_2} (\T^1 ) \, \d(\T^2 - \bar{\T}^1 ) \over (z_{12}^0)^{2
\D_1^g}}
\left( { z^0_{12} \over \bz_{12} } \right)^{2 \tD_{\a_1} } + \cO (k^{-2})
\eqno(8.7)$$
where we have used the fact that the two-point affine-Sugawara correlator
\linebreak
$A_g^{\a_1 \a_2} (z_0)=\eta^{\a_1 \a_2}(\T^1) \d(\T^2 -\bar{\T}^1)
/(z_{12}^0)^{2 \D_1^g} $
is proportional to the inverse carrier space metric $\eta^{\a_1 \a_2} (\T^1)$
of representation $\T^1$.

Choosing $n=3$ in eq.(8.1), we obtain the high-level three-point
correlators of ICFT,
$$
\bA^{\a_1\a_2\a_3} (\bz_1 \, z_1^0 \,\T^1,\bz_2 \,z_2^0 \, \T^2, \bz_3 \,z_3^0
\,\T^3)
 = A_g^{\a_1 \a_2 \a_3}
(z_1^0 \, \T^1,z_2^0 \, \T^2 ,z_3^0\, \T^3)
\;\;\;\;\;\;\;\;\;\;\;\;\;\;  \;\;\;\;\;
 \eqno(8.8)$$
$$
\ti
\left( { z^0_{12} \over \bz_{12} } \right)^{ \tD_{\a_1} +\tD_{\a_2}-\tD_{\a_3}}
\left( { z^0_{13} \over \bz_{13} } \right)^{ \tD_{\a_1} +\tD_{\a_3}-\tD_{\a_2}}
\left( { z^0_{23} \over \bz_{23} } \right)^{ \tD_{\a_2} +\tD_{\a_3}-\tD_{\a_1}}
 + \cO (k^{-2})
$$
where the three-point affine-Sugawara correlators $A_g^{\a_1 \a_2 \a_3}$ are
proportional to the Clebsch-Gordan coefficients of the decomposition
$\T^1 \oti \T^2 $ into $\bar{\T}^3$. The result
(8.8) is an independent confirmation of a conclusion reached in
Ref.\cite{WI2}: The
high-level fusion rules of the broken affine primaries
follow the Clebsch-Gordan coefficients of the representations.

\ni E. {\it Invariant four-point correlators}. The high-level invariant
four-point correlators of ICFT are known \cite{WI2}. To check our results
against
these, recall that the invariant correlators
were obtained in the KZ gauge, where
$$
 \bY^{\a} (\bu,u_0) = \left( \prod_{i<j}^4  \bz_{ij}^{\bc_{ij}}
\right) \bA^{\a} ( \bz,z_0)   \;\;\;\;,\;\;\;\;
 \bu = { \bz_{12} \bz_{34} \over \bz_{14} \bz_{32} }
\;\;\;,\;\;\; \a=(\a_1 \a_2 \a_3 \a_4)
\eqno(8.9a)$$
$$
 \eqalign{ \bc_{12} = \bc_{13}  = 0\;\;\;,\;\;\;\; \bc_{14} = 2 \,
\tD_{\a_1} \;\;\;&,\;\;\;\;
 \bc_{23} = \tD_{\a_1} + \tD_{\a_2} + \tD_{\a_3} - \tD_{\a_4} \cr
 \bc_{24} = -\tD_{\a_1} + \tD_{\a_2} - \tD_{\a_3} + \tD_{\a_4} \;\;\;
&,\;\;\;\;
 \bc_{34} = -\tD_{\a_1} - \tD_{\a_2} + \tD_{\a_3} + \tD_{\a_4} \;\;\;\;\; .
\cr}
\eqno(8.9b)$$
Substituting our result (8.1a) for $n=4$
into (8.9a), we find
$$
\bY^{\a} (\bu,u_0) = Y_{\tL}^{\a} (\bu,u_0) d(\a) \;\;\;\;,\;\;\;\;
d(\a) \equiv  \prod_{i <j}^4 (z_{ij}^0)^{\bc_{ij}}
\eqno(8.10)$$
after some algebra with the global Ward identity (7.2b).
Here, $d(\a)$ is an irrelevant constant and the $\bu$-dependent factor,
$$
Y_{\tL}^{\a} (\bu,u_0)  =  Y_g^{\b} (u_0)  (
\one +  {\tilde{P}_{ab} \over k} \left[
\T_a^1  \T_b^2 \ln \left( {\bu \over u_0} \right)
+\T_a^1  \T_b^3 \ln \left( {1-\bu \over 1-u_0} \right) \right]
 )_{\b}{}^{\a} + \cO (k^{-2})
\eqno(8.11a)$$
$$
Y_g(u_0) = A_g (u_0) \;\;\;\;,\;\;\;\;
\tL^{ab} = {\tilde{P}^{ab} \over 2k} + \cO (k^{-2})
\eqno(8.11b)$$
is exactly the result of Ref.\cite{WI2}.

\ni F.
 {\it Induced non-local conserved quantities}.
As seen for the coset correlators in Section 6,
 the  non-local conserved quantities of the high-level
biconformal correlators induce non-local conserved quantities
for the high-level conformal
correlators.

To see this, we need the induction identity
$$
\eqalign{
\left( \one + {P^{ab} \over k} \sum_{i <j} \T_a^i \T_b^j
\ln \left( {z_{ij} \over z_{ij}^0 } \right) \right)
& Q_a(\bz,z) \cr
=  Q_a(\bz,z_0)&  \left( \one + {P^{ab} \over k} \sum_{i <j}
\T_a^i
\T_b^j \ln \left( {z_{ij} \over z_{ij}^0 } \right)  \right) + \cO (k^{-2} )
\cr}
\eqno(8.12)$$
which follows from (7.1d) in analogy with eq.(6.5). Then, following the
argument in
(6.6) and (6.7), we find that
$$
\bA(\bz,z_0) Q_a(\bz,z_0) = \cO (k^{-2} )
\;\;\;\;,\;\; a= 1 , \ldots , {\rm dim}\, g \eqno(8.13a)$$
$$
\eqalignno{
Q_a(\bz,z_0) & = Q^g_a + [Q^g_a, {\tilde{P}^{ab} \over k} \sum_{i <j}
\T_a^i \T_b^j \ln \left( {\bz_{ij} \over z_{ij}^0 } \right)] + \cO (k^{-2})
&(8.13b) \cr
& = \sum_{i=1}^n \T_a^i + { i \tilde{P}^{b(c} f_{ab}{}^{d)} \over k}
\sum_{i < j} \T_c^i \T_d^j \ln \left( { \bz_{ij} \over z_{ij}^0 } \right)
+ \cO (k^{-2} )
&(8.13c) \cr } $$
is induced from the invariance (7.1c) of the biconformal correlators.
This result can also be verified directly from the high-level
correlators in (8.1).

Taken together with the induced non-local $g$-invariance (6.7) of the
finite-level coset correlators, the result (8.13) strongly suggests that
$$
Q_a (\bz,z_0) \;\;\;\;, \;\;\;\; Q_a(z_0,z)
\eqno(8.14)$$
are the
finite-level induced non-local  conserved quantities for the
$\tL$ and $L$ theory respectively. This conjecture should be considered in
parallel with the conjectured form of the induced flat connections in
eq.(8.5).

\ni G. {\it Physical singularities}. The high-level correlators
(8.1) exhibit the correct singularities in all channels
$\bz_i \ra \bz_j$. For simplicity, we discuss this result for $\bz_1 \ra
\bz_2$,
 called the 1-2 channel, but any pair of $\bz$'s may be similarly chosen.

To begin, we expand the affine-Sugawara correlator
$A_g^{\a}(z_0)$ in (8.1a) in a complete set of invariant tensors
$v_n^{\a}$ of
$\T^1 \oti \cdots \oti \T^n$,
$$
A_g^{\a}(z_0)= \sum_{r,\x,[R]} \cA_g^{(n)}(r,\x,[R];z_0) v_n^{\a} (r,\x,[R])
\eqno(8.15a)$$
$$
A_g(z_0) Q_a^g = v_n (r,\x,[R]) Q_a^g =0 \;\;\;\;,\;\;\;\;
Q_a^g = \sum_{i=1}^n \T_a^i
\eqno(8.15b)$$
where the coefficients
$\cA_g^{(n)}$ are related to the $n$-point affine-Sugawara blocks
at $z_0$. In (8.15), we have chosen the invariant tensors in a basis
appropriate to the 1-2 channel,
$$
v_n^{\a_1 \a_2 \ldots \a_n} (r, \xi,[R])
=\sum_{\a_r \a_{\br} } v_3^{\a_1 \a_2 \a_r} (r,\xi)
\, \eta_{\a_r \a_{\br} } (\T^r) \, v_{n-1}^{\a_{\br} \a_3 \ldots \a_n}([R])
\eqno(8.16a)$$
$$
 v_3  (\T_a^1 + \T_a^2 + \T_a^r) =0
\;\;\;\;,\;\;\;\;
v_{n-1} (\bar{\T}_a^r + \T_a^3 + \ldots + \T_a^n) = 0
\eqno(8.16b)$$
where $\T^r$ is an irrep of $g$ in the 1-2 channel, $\eta_{\a_r \a_{\br} }
(\T^r)$ is its carrier space metric, and $v_3^{\a}$, $v_{n-1}^{\a}$ are the
invariant tensors of
$$
\T^1 \oti \T^2 \oti \T^r \;\;\;\;,\;\;\;\;
\bar{\T}^r \oti \T^3 \oti \cdots \oti \T^n
\eqno(8.17)$$
respectively. The argument
$\x$ in (8.15), (8.16a) labels different copies of $\T^r$ and $[R]$
collects the other couplings of $v_{n-1}$.
The result (8.16) was obtained by studying Haar integration
over $n$ matrix representations $g(\T^i),\,i =1, \ldots ,n$ of the group $g$.

It will also be helpful to recall that \cite{WI2}
$$
v_3^{\a_i\a_j \a_r }(r,\x) =
\sum_{\a_{\br}} \cl{\a_i}{\a_j}{\br(\x)}{i}{j}{\a_{\br}} \et^{\a_{\br}\a_{r}}
(\T^r)
\eqno(8.18a)$$
$$
 v_3^{\b_1 \b_2 \a_r} (r,\x)
[2 \tL^{ab} \T_a^1  \T_b^2]_{\b_1 \b_2 }{}^{\a_1 \a_2}
= v_3^{\a_1 \a_2 \a_{r}} (r,\x)
 (\tD_{\a_r} (\T^r) - \tD_{\a_1}(\T^1) - \tD_{\a_2}(\T^2))
\eqno(8.18b)$$
where $(\cdots )$ is the Clebsch-Gordan coefficient of $\T^i \oti \T^j$ into
$\bar{\T}^r $.

Using (8.15), (8.16) and (8.18b), we extract the form of
the high-level
correlators as $\bz_1 \ra \bz_2$,
$$
\bA^{\a}(\bz,z_0) \smash{ \mathop{\simeq} \limits_{\bz_1  \ra \bz_2} }
\sum_{{r, \xi, [R] \atop \a_r \a_{\br}} }
\cA_g (r,\xi,[R];z_0)
v_3^{\a_1 \a_2 \a_r} (r,\xi) \, \eta_{\a_r \a_{\br}} (\T^r )
\left( { \bz_{12} \over z_{12}^0 }\right)^{\tD_{\a_r} - \tD_{\a_1} -
\tD_{\a_2}}
$$
$$
\times
A_{(n-1)}^{\a_{\br} \a_3 \ldots \a_n} (\bz_2\,z_2^0\,\bar{\T}^r,
\bz_3\,z_3^0\,\T^3 , \ldots,
\bz_n\,z_n^0 \,\T^n) [R]
+ \cO (k^{-2})
\eqno(8.19a)$$
$$
\eqalign{
A_{(n-1)}^{\a_{\br} \a_3 \ldots \a_n} & (\bz_2 \, z_2^0 \,\bar{\T}^r,
\bz_3\,z_3^0\,\T^3 , \ldots,
\bz_n\,z_n^0 \,\T^n) [R] \cr
=   v_{n-1}^{\b_{\br} \b_3}  & {}^{ \ldots \b_n}([R])
[ 1 + { \tilde{P}^{ab} \over k} ( \sum_{j=3}^n \bar{\T}^r_a \T_b^j
\ln \left( {\bz_{2j} \over z_{2j}^0}\right) +
\sum_{3 \leq i < j}^n \T_a^i \T_b^j \ln \left(
{\bz_{ij} \over z_{ij}^0 } \right) )  ]_{\b_{\br}
\b_3 \ldots \b_n}^{\;\;\a_{\br} \a_3 \ldots \a_n}
\;\;. \cr } \eqno(8.19b)$$
This result shows the correct factorization of the high-level $n$-point
correlators into
the high-level three-point correlators times the high-level
$(n-1)$-point correlators, including the correct conformal weight factor for
broken affine primaries in the 1-2 channel.
Since the three-point invariant tensors (8.18a) are proportional to
the Clebsch-Gordan coefficients, the high-level fusion rules \cite{WI2} of ICFT
are also visible here.

\section{Conclusions}

Irrational conformal field theory (ICFT) includes rational conformal
field theory as a small subspace, and the affine-Virasoro Ward identities
describe the biconformal correlators of ICFT.

We have seen that the Ward identities are equivalent to a linear
partial differential system with flat connections and unsuspected new non-local
conserved quantities. In this formulation, the equations of ICFT are clearly
seen as generalized KZ equations, including the KZ equation itself as the
simplest example.

As examples of the new formulation we solved for the coset correlators, the
correlators of the affine-Sugawara nests and the high-level $n$-point
correlators of ICFT. In the latter case, our
results agree with the known high-level
four-point correlators and continue to show good physical behavior for $n \neq
4$.

Having seen this elegant structure in ICFT, we may remark on an outstanding
technical problem of the program: We have been studying ICFT primarily by
pedestrian methods, that is, by computing and summing the connection moments.
It is important now to develop new methods which compute the flat connections
more directly.

\section*{Acknowledgments}

We thank M. Douglas for suggesting this investigation.

We also thank the Niels Bohr Institute and the CERN Theory Division for
their hospitality during part of this work. NO also thanks the Theoretical
High Energy Physics Group at Nijmegen University for their hospitality.

The work of MBH was supported in part by the Director, Office of
Energy Research, Office of High Energy and Nuclear Physics, Division of
High Energy Physics of the U.S. Department of Energy under Contract
DE-AC03-76SF00098 and in part by the National Science Foundation under
grant PHY90-21139. The work of NO was supported in part by a
S\'ejour Scientifique de Longue Dur\'ee of the Minist\`ere des Affaires
Etrang\`eres.

\bigskip
\bigskip
\centerline{\bf Appendix A: Invariant flat connections}
\bigskip

Following the development in Sections 2 and 3, we may also translate the
invariant four-point Ward identities [13,14]
$$ \bpa^q \pa^p Y(\bu,u) \ve_{\bu = u} = Y_g (u) W_{qp} (u)
\eqno(A.1a) $$
$$ Y_g(u) = Y(u,u) \;\;\;\;, \;\;\;\; Y_g (u) Q^g_a =0 \;\;\;\;,\;\;\;\;
Q^g_a = \sum_{i=1}^4 \T_a^i
\eqno(A.1b) $$
into the system of invariant flat connections,
$$ \bpa Y (\bu,u) = Y (\bu,u)  \bW (\bu,u)  \;\;\;\;,\;\;\;\;
\pa Y (\bu,u)  = Y (\bu,u) W (\bu,u)
\eqno(A.2a) $$
$$ Y(\bu,u) Q_a (\bu,u) =0 \;\;\;\;, \;\; a=1 , \ldots, {\rm dim}\,g
\eqno(A.2b) $$
where $Y_g(u)$ is the invariant affine-Sugawara correlator, $Y(\bu,u)$ is the
invariant four-point biconformal correlator  and
$Q_a (\bu,u) $ are the invariant non-local conserved generators of Lie $g$.

Drawing on known results [13,14] for the invariant system, we
obtain the following partial list of
relevant relations,
$$
(\pa + W) \bW = (\bpa + \bW) W \eqno(A.3a) $$
$$ W_{qp} = (\bpa + \bW )^q (\pa + W)^p \one \ve_{\bu = u}
\eqno(A.3b) $$
$$ Y(\bu,u) = Y_g (u) \Fb (\bu,u) = Y_g(\bu) F (\bu,u)
\eqno(A.3c) $$
$$
Y_g(u) = Y_g (u_0) Y(u,u_0) \;\;\;\;,\;\;\;\;
\pa_u Y_g (u,u_0) = Y_g(u,u_0) W^g(u) \;\;\;\;,\;\;\;\; Y_g(u_0,u_0) =\one
\eqno(A.3d) $$
$$ \Fb(\bu,u) \equiv U^* \mbox{e}^{\int_{u}^{\bu} {\rm d} \bu ' \bW(\bu ',u) }
= \sum_{q=0}^{\infty}
 { (\bu-u)^q \over q!} W_{q0}(u) \eqno(A.3e) $$
$$ F(\bu,u) \equiv U^* \mbox{e}^{\int_{\bu}^u {\rm d} u ' W(\bu,u') }
= \sum_{p=0}^{\infty}
  { (u-\bu)^p \over p!}  W_{0p} (\bu) \eqno(A.3f) $$
$$ \Fb (\bu,u)  = Y_g (\bu,u)  F (\bu,u)  \eqno(A.3g) $$
$$ \bW = \Fb^{-1} \bpa \Fb = F^{-1} (\bpa + W^g(\bu) ) F
\eqno(A.3h) $$
$$ W = F^{-1} \pa F = \Fb^{-1} (\pa + W^g(u) ) \Fb
\eqno(A.3i) $$
$$ \bW (u,u) = W_{10}(u) = 2\tL^{ab} \left( { \T_a^1 \T_b^2  \over u}
+ { \T_a^1 \T_b^3  \over u-1} \right) \eqno(A.3j) $$
$$ W (u,u) = W_{01}(u) = 2L^{ab} \left( { \T_a^1 \T_b^2  \over u}
+ { \T_a^1 \T_b^3  \over u-1} \right) \eqno(A.3k) $$
$$ \bW (u,u) + W(u,u) =  W^g(u) = 2L_g^{ab} \left( { \T_a^1 \T_b^2  \over u}
+ { \T_a^1 \T_b^3  \over u-1} \right) \eqno(A.3l) $$
$$ Q_a (\bu,u) = F^{-1}(\bu,u) Q^g_a F(\bu,u) = \Fb^{-1}(\bu,u)Q^g_a \Fb(\bu,u)
\eqno(A.3m) $$
$$
[Q_a (\bu,u) , Q_b (\bu,u) ] = i f_{ab}{}^c Q_c(\bu,u)
\eqno(A.3n) $$
$$
\bar{D} Q_a \equiv \bpa Q_a + [\bW,Q_a] =0 \;\;\;\;,\;\;\;\;
 D Q_a \equiv  \pa Q_a + [W,Q_a] = 0
\eqno(A.3o) $$
$$
\bar{D}( Y Q_a) = \bpa (Y Q_a) - (Y Q_a)  \bW =0 \;\;\;\;,\;\;\;\;
 D ( Y Q_a ) =  \pa (Y Q_a)  - (Y Q_a)  W = 0
\eqno(A.3p) $$
$$ Y_g (u) Q^g_a =0 \lra Y(\bu_0,u_0) Q_a(\bu_0,u_0) =0 \lra
Y(\bu,u) Q_a(\bu,u) =0
\eqno(A.3q) $$
in complete analogy with the results of Section 2 and 3.
In (A.3e) and (A.3f),
 $U^*$ is
anti-ordering in $u$.

For K-conjugation and crossing symmetry we obtain the relations,
$$ \Fb (\tL ,\bu,u) \;\;\;\;, \;\;\;\; F(L,\bu,u)
\eqno(A.4a) $$
$$
\bW (\tL,\bu,u) \;\;\;\;, \;\;\;\; W(L,\bu,u)
\eqno(A.4b) $$
$$ \Fb (\tL =L_*,\bu,u) = F(L=L_* ,u,\bu)
\;\;\;\;,\;\;\;\;
F (L =L_*,\bu,u) = \Fb (\tL=L_* ,u,\bu)
\eqno(A.4c) $$
$$
\bW (\tL= L_* ,\bu,u) = W (L = L_*,u,\bu)
\;\;\;\;,\;\;\;\;
W (L=L_*,\bu,u) = \bW (\tL = L_*,u,\bu)
\eqno(A.4d) $$
$$ Y(\bu,u) \ve_{\tL \lra L} = Y(u,\bu)
\eqno(A.4e) $$
$$ Q_a(\bu,u) \ve_{\tL \lra L} = Q_a(u,\bu)
\eqno(A.4f) $$
$$ \Fb (1-\bu,1-u) =  P_{23} \Fb (\bu,u) P_{23} \;\;\;\;,\;\;\;\;
 F (1-\bu,1-u) = P_{23} F(\bu,u) P_{23}
\eqno(A.4g) $$
$$
\bW (1-\bu,1-u) =- P_{23} \bW (\bu,u) P_{23} \;\;\;\;, \;\;\;\;
 W (1-\bu,1-u) = - P_{23} W (\bu,u) P_{23}
\eqno(A.4h) $$
$$
 Q_a(1-\bu,1-u) = P_{23} Q_a (\bu,u) P_{23}
\eqno(A.4i) $$
where $P_{23}$
is the exchange operator
which satisfies $P_{23} \T^2 P_{23} = \T^3,\; P_{23}^2 =1$.

For the general nest, we find the invariant evolution operators,
$$
\Fb [L=L_{g/h_1 / \ldots / h_{2n+1}} , \bu,u ]
= Y_g (\bu,u) Y_{h_1} (u,\bu) \cdots Y_{h_{2n+1}} (u,\bu)
\eqno(A.5a) $$
$$
F [L=L_{h_1 / \ldots / h_{2n+1}} , \bu,u ]
=  Y_{h_1} (u,\bu) \cdots Y_{h_{2n+1}} (u,\bu)
\eqno(A.5b) $$
$$
\Fb [L=L_{g/h_1 / \ldots / h_{2n}} , \bu,u ]
= Y_g (\bu,u) Y_{h_1} (u,\bu) \cdots Y_{h_{2n}} (\bu,u)
\eqno(A.5c) $$
$$
F [L=L_{h_1 / \ldots / h_{2n}} , \bu,u ]
=  Y_{h_1} (u,\bu) \cdots Y_{h_{2n}} (\bu,u)
\eqno(A.5d) $$
where $Y_{h_i} (u,u_0)$ is the invariant evolution operator of $h_i$,
$$
\pa Y_{h_i} (u,u_0) =Y_{h_i} (u,u_0) W^{h_i} (u) \;\;\;\;,\;\;\;\;
Y_{h_i} (u_0,u_0) = \one \;\;\;\;.
\eqno(A.6) $$
Using (A.3c,h,i) and (A.5), one may obtain the
invariant flat connections $\bW,\,W$
of the nests and their known \cite{WI2} invariant
four-point biconformal correlators.
We have also checked that $\bW (\bu,u_0)$ and $Q_a(\bu,u_0)$ are the induced
connections and induced non-local conserved generators of $g$ for the
invariant coset correlators $(Y_g(u_0) Y_h^{-1}(u,u_0))^{\a}$ obtained in
Ref.\cite{WI}.

The relevant high-level results are:
$$
 \Fb(\bu,u) = \one + { \tilde{P}^{ab} \over k} \left(
\T_a^1 \T_b^2 \ln \left( { \bu \over u} \right)
+ \T_a^1 \T_b^3 \ln \left( {1-\bu \over 1-u} \right)  \right)
   + \cO (k^{-2} ) \eqno(A.7a) $$
$$
 F(\bu,u) = \one + { P^{ab} \over k} \left( \T_a^1 \T_b^2
\ln \left( { u \over \bu} \right)
+ \T_a^1 \T_b^3 \ln \left( {1-u \over 1-\bu} \right)  \right)
   + \cO (k^{-2} ) \eqno(A.7b) $$
 $$ \bW(\bu,u) = W_{10} (\bu) + \cO (k^{-2}) \eqno(A.7c) $$
 $$ W(\bu,u) = W_{01} (u) + \cO (k^{-2}) \eqno(A.7d) $$
$$
\eqalign{
Y(\bu,u) = Y_g(u_0) & \left[ \one +
  { \tilde{P}^{ab} \over k} \left(
 \T_a^1 \T_b^2 \ln  \left( { \bu \over u_0} \right)
+ \T_a^1 \T_b^3 \ln \left( {1-\bu \over 1-u_0} \right)  \right)
  \right]  \cr
 \ti & \left[ \one + {P^{ab} \over k} \left(
\T_a^1 \T_b^2 \ln \left( { u \over u_0} \right)
+ \T_a^1 \T_b^3 \ln \left( {1-u \over 1-u_0} \right)  \right) \right]
   + \cO (k^{-2} )  \cr}
\eqno(A.7e) $$
$$ Y (\bu,u) Q_a^g (\bu,u) = \cO (k^{-2})
\eqno(A.7f) $$
$$
Q_a (\bu,u) = Q_a^g + [Q_a^g, {P^{ab} \over k} (
\T_a^1 \T_a^2 \ln \left( {u \over \bu} \right) +
\T_a^1 \T_b^3 \ln \left( {1-u \over 1-\bu} \right)) ] + \cO (k^{-2} )
\eqno(A.7g) $$
$$  = Q_a^g + [Q_a^g, {\tilde{P}^{ab} \over k} (
\T_a^1 \T_a^2 \ln \left( {\bu \over u} \right) +
\T_a^1 \T_b^3 \ln \left( {1-\bu \over 1-u} \right)) ] + \cO (k^{-2} )
\eqno(A.7h) $$
$$ Y_{\tL} (\bu,u_0) Q_a^g (\bu,u_0) = \cO (k^{-2} ) \;\;\;\; .
\eqno(A.7i) $$
The high-level invariant conformal correlators $Y_{\tL} (\bu,u_0)$ in
 (8.11a) and (A.7i) can be read from the factorized form of the
high-level invariant
 biconformal correlators in (A.7e).
We have also checked that $\bW (\bu,u_0)$ and $Q_a(\bu,u_0)$ are the induced
connection and induced non-local conserved generators of $g$ for the
high-level correlators $Y_{\tL} (\bu,u_0) $.

\bigskip
\bigskip
\centerline{\bf Appendix B: Flat connections and connection moments}
\bigskip
In this appendix, we work out in further detail the formula of the text,
$$ W_i (\bz,z) = F^{-1} (\bz,z) \pa_i F(\bz,z)
\eqno(B.1a) $$
$$ F(\bz,z) =\sum_{p=0}^{\infty} {1 \over p!}
\sum_{i_1 \ldots i_p} \prod_{\m=1}^p (z_{i_{\m}} - \bz_{i_{\m}} )
W_{0,i_1 \ldots i_p} (\bz)
\eqno(B.1b) $$
which expresses the flat connection $W_i$ in terms of its one-sided
connection moments $W_{0,i_1 \ldots i_p}$.

Using $ (1 + x)^{-1} = \sum_{r=0}^{\infty} (-)^r x^r$, we have the more
explicit form of the flat connection,
$$ W_i (\bz,z) =
\left[ \sum_{r=0}^{\infty} (-)^r
 \left( \sum_{p=1}^{\infty} {1 \over p!} \sum_{i_1 \ldots i_p}
\prod_{\m =1}^p (z_{i_{\m}} - \bz_{i_{\m}} ) W_{0, i_1 \ldots i_p} (\bz)
\right)^r
\right] $$
$$ \times \left(
\sum_{s=0}^{\infty} {1 \over s!} \sum_{k_1 \ldots k_s}
\prod_{\n =1}^s (z_{k_{\n}} - \bz_{k_{\n}} ) W_{0, k_1 \ldots k_s i} (\bz)
\right)
\eqno(B.2) $$
which can be rearranged as follows,
$$ W_i (\bz,z) = \sum_{p=0}^{\infty} {1 \over p!} \sum_{i_1 \ldots i_p}
\prod_{\m = 1}^p (z_{i_{\m}} - \bz_{i_{\m}} )  \r_{ i_1 \ldots i_p i} (\bz)
\eqno(B.3a) $$
$$ \eqalignno{
\r_{ i_1 \ldots i_pi} (\bz) & \equiv \pa_{i_1} \cdots \pa_{i_p} W_i (\bz,z)
\ve_{z = \bz} &(B.3b) \cr
  = \sum_{r=0}^{\infty} & (-)^r \sum_{p_1=1}^{\infty}
 \cdots \sum_{p_r =1}^{\infty} \sum_{s=0}^{\infty}
\d(p - s - \sum_{i=1}^r p_i ) {1 \over s! \prod_{i=1}^r p_i ! }
& \cr
 \ti \sum_{\P(i_1 \ldots i_p)} &  W_{0,i_1 \ldots i_{p_1}}
W_{0, i_{p_1 +1} \ldots i_{p_1 + p_2}} \cdots
W_{0,i_{p-p_r-s+1} \ldots i_{p-s} }
W_{0, i_{p-s+1} \ldots i_p i}\;\;. &(B.3c) \cr}
 $$
This is a computationally convenient form of the flat connection, and we find,
for example, that
$$ \r_i = W_{0,i} \eqno(B.4a) $$
$$ \r_{ i_1i} = W_{0, i_1 i} - W_{0,i_1} W_{0,i} \eqno(B.4b) $$
$$ \eqalign{
\r_{ i_1 i_2i} & = W_{0, i_1 i_2 i} - W_{0,(i_1} W_{0,i_2)i} - W_{0,i_1 i_2}
W_{0,i} + W_{0,(i_1} W_{0,i_2)}
 W_{0,i}   \cr
& = \left[ \frac{1}{2} W_{0, i_1 i_2 i} - W_{0,i_1} W_{0,i_2i} - \frac{1}{2}
W_{0,i_1 i_2}
W_{0,i} + W_{0,i_1} W_{0,i_2}
 W_{0,i}\right] + ( i_1 \lra i_2)  \;\;\;\;. \cr} \eqno(B.4c)  $$
Similarly, we have the alternate expansion,
$$ W_i (\bz,z) = \sum_{p=0}^{\infty} {1 \over p!} \sum_{i_1 \ldots i_p}
\prod_{\m = 1}^p(\bz_{i_{\m}} - z_{i_{\m}} ) \s_{ i_1 \ldots i_pi} (z)
\eqno(B.5a) $$
$$
\eqalign{
 \s_{ i_1 \ldots i_pi} (z) &  \equiv \bpa_{i_1} \cdots \bpa_{i_p} W_i (\bz,z)
\ve_{\bz = z}  \cr
&  = \sum_{r=0}^p { (-)^{p-r} \over (p-r)! r! } \sum_{\P (i_1 \ldots i_p)}
\pa_{i_1} \cdots \pa_{i_r} \r_{ i_{r+1} \ldots i_p i}(z)
\cr}
\eqno(B.5b) $$
$$
\s_i = W_{0,i} \eqno(B.5c) $$
$$ \s_{ i_1i} = \pa_{i_1} W_{0,i} -W_{0,i i_1} + W_{0,i_1} W_{0,i}
\eqno(B.5d) $$
$$ \eqalign{
\;\;\;\;\;\;\; \s_{ i_1 i_2i} = & \pa_{i_1} \pa_{i_2} W_{0,i}
+ \pa_{(i_1} W_{0,i_2)} W_{0,i}
  - \pa_{(i_1} W_{0,i_2)i}
 + W_{0, i_i i_2 i}
\cr  & - W_{0,(i_1} W_{0,i_2)i} - W_{0,i_1 i_2}
W_{0,i} + W_{0,(i_1} W_{0,i_2)}
 W_{0,i}
\;\;\;\;. \cr} \eqno(B.5e) $$
Using (2.17a), similar results for $\bW_i = \Fb^{-1} \bpa_i \Fb$
are easily obtained.

Using eqs.(A.3f) and (A.3i) of Appendix A, we also give the
corresponding results for the invariant flat connection
and its one-sided connection moments:
$$ W (\bu,u) =
\left[ \sum_{r=0}^{\infty} (-)^r
 \left( \sum_{p=1}^{\infty}
 {(u - \bu )^p \over p!} W_{0p} (\bu) \right)^r
\right]
 \left(
\sum_{s=0}^{\infty}
 {(u - \bu )^s \over s!} W_{0,s+1} (\bu) \right)
\eqno(B.6a) $$
$$ = \sum_{p=0}^{\infty}
{ (u  - \bu  )^p \over p!}  \r_{p} (\bu)
= \sum_{p=0}^{\infty}
{(\bu -u  )^p \over p!} \s_{p} (u)
\eqno(B.6b) $$
$$
\r_{p} (\bu)  \equiv \pa^p  W (\bu,u)
\ve_{u = \bu}  \;\;\;\;, \;\;\;\;
 \s_{p} (u)   \equiv \bpa^p  W (\bu,u)
\ve_{\bu = u}
  = \sum_{r=0}^p  (-)^{p-r} \ch{p}{r}
\pa^r \r_{p-r }(u)
\eqno(B.6c) $$
$$ \r_p (\bu) = \sum_{r=0}^{\infty}  (-)^r \sum_{p_1=1}^{\infty}
 \cdots \sum_{p_r =1}^{\infty} \sum_{s=0}^{\infty}
\d(p - s - \sum_{i=1}^r p_i ) {p! \over s! \prod_{i=1}^r p_i ! }
  W_{0p_1}
W_{0 p_2} \cdots
W_{0p_r}
W_{0,s+1}
\eqno(B.6d)  $$
$$ \r_{0} = W_{01} \eqno(B.6e) $$
$$ \r_{1} = W_{02} - W_{01}^2  \eqno(B.6f) $$
$$ \r_{2} = W_{03} - 2 W_{01} W_{02} - W_{02}
W_{01} + 2 W_{01}^3
  \eqno(B.6g) $$
$$
\s_0 = W_{01} \eqno(B.6h) $$
$$ \s_{1} =\pa W_{01} -W_{02} + W_{01}^2
\eqno(B.6i) $$
$$ \s_{2} =
  \pa^2 W_{01} +2 \pa W_{01}^2 - 2\pa W_{02}
 + W_{03} - 2 W_{01} W_{02} - W_{02}
W_{01} +2 W_{01}^3
\;\;\;\;. \eqno(B.6j) $$

\bigskip
\bigskip
\centerline{\bf Appendix C: Non-local generators of ${\bf g}$ and connection
moments}
\bigskip

In this appendix, we work out in further detail the formula of the text,
$$ Q_a (\bz,z) = F^{-1} (\bz,z) Q_a^g F(\bz,z) \;\;\;\;, \;\;
a= 1, \ldots, {\rm dim}\,g
\eqno(C.1) $$
which expresses the non-local generators of $g$ in terms of the  global
generators of $g$ and the one-sided
connection moments $W_{o,i_1 \ldots i_p}$.

Following the development of Appendix B, we find
$$ \eqalign{
Q_a (\bz,z) & = \sum_{p=0}^{\infty} {1 \over p!} \sum_{i_1 \ldots i_p}
\prod_{\m = 1}^p (z_{i_{\m}} - \bz_{i_{\m}} )  (q_a)_{ i_1 \ldots i_p } (\bz)
\cr & = \sum_{p=0}^{\infty} {1 \over p!} \sum_{i_1 \ldots i_p}
\prod_{\m = 1}^p (\bz_{i_{\m}} - z_{i_{\m}} )  (\bq_a)_{ i_1 \ldots i_p } (z)
\cr} \eqno(C.2a) $$
$$ (q_a)_{ i_1 \ldots i_p} (\bz)  \equiv \pa_{i_1} \cdots \pa_{i_p} Q_a (\bz,z)
\ve_{z = \bz} \;\;\;\;,\;\;\;\;
 (\bq_a)_{ i_1 \ldots i_p} (z)  \equiv \bpa_{i_1} \cdots \bpa_{i_p} Q_a (\bz,z)
\ve_{\bz = z}
\eqno(C.2b) $$
$$ \eqalign{
(q_a)_{ i_1 \ldots i_p} (\bz)
  & = \sum_{r=0}^{\infty}  (-)^r \sum_{p_1=1}^{\infty}
 \cdots \sum_{p_r =1}^{\infty} \sum_{s=0}^{\infty}
\d(p - s - \sum_{i=1}^r p_i ) {1 \over s! \prod_{i=1}^r p_i ! }
 \cr
 \ti & \sum_{\P(i_1 \ldots i_p)}   W_{0,i_1 \ldots i_{p_1}}
W_{0, i_{p_1 +1} \ldots i_{p_1 + p_2}} \cdots
W_{0,i_{p-p_r-s+1} \ldots i_{p-s} }
Q_a^g \, W_{0, i_{p-s+1} \ldots i_p } \cr}
 \eqno(C.2c) $$
$$ (q_a) = Q_a^g \eqno(C.2d) $$
$$ (q_a)_{ i_1} = [ Q_a^g, W_{0, i_1 }]  \eqno(C.2e) $$
$$
(q_a)_{ i_1 i_2}  =[Q_a^g, W_{0, i_1 i_2 }]  + W_{0,(i_1} [ W_{0,i_2)}, Q_a^g]
 \eqno(C.2f)  $$
$$
(\bq_a) = Q_a^g
\eqno(C.2g) $$
$$ (\bq_a)_{ i_1} = -[Q_a^g ,W_{0,i_1 }]
\eqno(C.2h) $$
$$
(\bq_a)_{ i_1 i_2} =
 \pa_{(i_1} [W_{0,i_2)}, Q_a^g]
 + [Q_a^g,W_{0, i_i i_2 } ]
+ W_{0,(i_1} [ W_{0,i_2)}, Q_a^g ]   \;\;\;\;.
 \eqno(C.2i) $$
For the invariant non-local conserved quantities $Q_a(\bu,u) = F^{-1} (\bu,u)
Q_a^g F(\bu,u)$ in (A.3m), we find the analogous results,
$$
Q_a (\bu,u)  = \sum_{p=0}^{\infty}
{ (u-\bu)^p \over p!}   (q_a)_{ p } (\bu)
 = \sum_{p=0}^{\infty}
{ (\bu-u)^p \over p!}   (\bq_a)_{ p } (u)
 \eqno(C.3a) $$
$$ (q_a)_{ p} (\bu)  \equiv \pa^p  Q_a (\bu,u)
\ve_{u = \bu} \;\;\;\;,\;\;\;\;
 (\bq_a)_{p} (u)  \equiv \bpa^p  Q_a (\bu,u)
\ve_{\bu = u}
\eqno(C.3b) $$
$$ \eqalign{
(q_a)_{ p} (\bu)
   = \sum_{r=0}^{\infty}  (-)^r & \sum_{p_1=1}^{\infty}
 \cdots \sum_{p_r =1}^{\infty} \sum_{s=0}^{\infty}
\d(p - s - \sum_{i=1}^r p_i ) {p!  \over s! \prod_{i=1}^r p_i ! }
\cr & \ti     W_{0p_1}
W_{0  p_2} \cdots
W_{0p_r }
Q_a^g \, W_{0,s+1 }
\cr}  \eqno(C.3c) $$
$$ (q_a)_0 = Q_a^g \eqno(C.3d) $$
$$ (q_a)_{ 1} = [ Q_a^g, W_{01 }]  \eqno(C.3e) $$
$$
(q_a)_{ 2}  =[Q_a^g, W_{02 }]  + 2 W_{01} [ W_{01}, Q_a^g]
 \eqno(C.3f)  $$
$$
(\bq_a)_0 = Q_a^g
\eqno(C.3g) $$
$$ (\bq_a)_{ 1} = -[Q_a^g ,W_{01 }]
\eqno(C.3h) $$
$$
(\bq_a)_{ 2} =
 2 \pa [W_{01}, Q_a^g]
 + [Q_a^g,W_{02 } ]
+ 2 W_{01} [ W_{01}, Q_a^g ]   \;\;\;\;.
 \eqno(C.3i) $$

\end{document}